\documentclass[aps,prc,twocolumn,showpacs,preprintnumbers,nofootinbib,float,superscriptaddress]{revtex4-1}
\usepackage{epsfig}
\usepackage{color}
\usepackage[dvipsnames]{xcolor}
\usepackage{float,amsmath,amssymb}
\usepackage{graphicx}
\usepackage{url}
\usepackage{bbold}
\usepackage[utf8]{inputenc}
\usepackage{physics}
\newcommand{\by}{\mathbf{y}}

\newcommand{\xpom}{x_\mathbb{P}}

\newcommand{\gev}{\mathrm{GeV}}

\newcommand{\rt}{{\mathbf{r}_\perp}}

\newcommand{\xt}{{\mathbf{x}_\perp}}

\newcommand{\yt}{{\mathbf{y}_\perp}}
\newcommand{\bt}{{\mathbf{b}_\perp}}
\newcommand{\bti}{{\mathbf{b}_{\perp,i}}}
\newcommand{\Deltat}{{\boldsymbol{\Delta}_\perp}}

\newcommand{\nc}{{N_\mathrm{c}}}
\newcommand{\jpsi}{$\mathrm{J}/\psi$ }

\newcommand{\btheta}{{\boldsymbol{\theta}}}
\newcommand{\CalP}{{\mathcal{P}}}
\usepackage[breaklinks,colorlinks,citecolor=citcolor,urlcolor=blue,linkcolor=lcolor]{hyperref}
\definecolor{lcolor}{rgb}{0.5,0,0}
\definecolor{citcolor}{rgb}{0,0.3,0.0}

\begin{document}

\title{Bayesian inference of the fluctuating proton shape}

\author{Heikki M\"antysaari}
\affiliation{Department of Physics, University of Jyv\"askyl\"a, P.O. Box 35, 40014 University of Jyv\"askyl\"a, Finland}
\affiliation{Helsinki Institute of Physics, P.O. Box 64, 00014 University of Helsinki, Finland}

\author{Bj\"orn Schenke}
\affiliation{Physics Department, Brookhaven National Laboratory, Upton, NY 11973, USA}

\author{Chun Shen}
\affiliation{Department of Physics and Astronomy, Wayne State University, Detroit, Michigan 48201, USA}
\affiliation{RIKEN BNL Research Center, Brookhaven National Laboratory, Upton, NY 11973, USA}

\author{Wenbin Zhao}
\affiliation{Department of Physics and Astronomy, Wayne State University, Detroit, Michigan 48201, USA}

\begin{abstract}
Using Bayesian inference, we determine probabilistic constraints on the parameters describing the fluctuating structure of protons at high energy. We employ the color glass condensate framework supplemented with a model for the spatial structure of the proton, along with experimental data from the ZEUS and H1 Collaborations on coherent and incoherent diffractive \jpsi production in e+p collisions at HERA. This data is found to constrain most model parameters well. This work sets the stage for future global analyses, including experimental data from e+p, p+p, and p+A collisions, to constrain the fluctuating structure of nucleons along with properties of the final state.
\end{abstract}

%\pacs{11.15Bt, 04.25.Nx, 11.10Wx, 12.38Mh}
\maketitle

%%%%%%%%%%%%%%%%%%%%%%%%%%%%%%%%%%%%%%%%%%%%%%%%%%%%%%%%%%%%%%%

\section{Introduction}

Extracting the multi-dimensional structure of protons and nuclei is one of the main goals of future Deep Inelastic Scattering (DIS) facilities such as the Electron-Ion Collider~\cite{AbdulKhalek:2021gbh,Aschenauer:2017jsk}, LHeC/FC-he~\cite{Agostini:2020fmq} and EicC~\cite{Anderle:2021wcy}. Exclusive processes like \jpsi production are especially powerful probes of the hadron structure at a small longitudinal momentum fraction $x$ for two reasons. First, the exclusive nature of the process requires at lowest order at least two gluons to be exchanged with the target, rendering the cross section approximately proportional to the squared gluon distribution~\cite{Ryskin:1992ui}. Additionally, only in exclusive processes is it possible to measure the total momentum transfer, which is Fourier conjugate to the impact parameter and thereby provides access to the transverse geometry.

Understanding the proton structure, including its event-by-event fluctuations~\cite{Mantysaari:2020axf}, is of fundamental interest. Additionally, knowledge of the spatial structure of the colliding objects in hadronic and heavy-ion collisions is required in order to construct realistic initial conditions that can be coupled to relativistic hydrodynamic simulations to describe the space-time evolution of the produced Quark-Gluon Plasma (QGP). 
Besides heavy-ion collisions, collective phenomena that can be interpreted as signatures of QGP production have been seen in small systems such as proton/deuteron/$^3$He - nucleus \cite{ALICE:2012eyl, CMS:2012qk, ATLAS:2012cix,PHENIX:2018lia,STAR:2019zaf}, proton-proton \cite{CMS:2010ifv}, and even photon-nucleus collisions \cite{ATLAS:2021jhn}, see \cite{Dusling:2015gta,Loizides:2016tew,Schlichting:2016sqo,Nagle:2018nvi,Schenke:2019pmk,Schenke:2021mxx} for reviews. 
In such small collision systems with a proton projectile, the detailed fluctuation spectrum of the proton geometry is particularly important to determine if QGP is indeed produced.

It is possible to constrain the proton structure from hadronic collisions by performing a statistical analysis to extract both the transport coefficients describing the matter produced in proton-lead collisions (see e.g. Refs.~\cite{JETSCAPE:2021ehl,JETSCAPE:2020mzn,JETSCAPE:2020shq,Bernhard:2019bmu,Bernhard:2016tnd,Parkkila:2021yha,Parkkila:2021tqq}), as well as the proton's fluctuating geometry, by comparing with the LHC data as in Ref.~\cite{Moreland:2018gsh}. Another approach, which we take in this work, is to use exclusive DIS data from HERA, especially exclusive vector meson production~\cite{H1:2013okq,Aktas:2005xu,Chekanov:2002xi,Chekanov:2002rm,Aktas:2003zi}, 
as a complementary input to constrain the proton shape fluctuations, as initially suggested in Ref.~\cite{Mantysaari:2016ykx}. In the future, the Electron-Ion Collider will provide a vast amount of precise vector meson production data with proton and nuclear targets that will provide further constraints on e.g.~momentum fraction $\xpom$ and the nuclear mass number $A$ dependence. Additionally, Ultra Peripheral Collisions~\cite{Klein:2019qfb,Bertulani:2005ru} at RHIC~\cite{PHENIX:2009xtn} and at the LHC~\cite{ALICE:2014eof,ALICE:2018oyo,LHCb:2014acg, LHCb:2018rcm,CMS:2016itn,ALICE:2021tyx} provide access to very high energy photoproduction processes and to effects of a nuclear environment on nucleon substructure fluctuations at high energy~\cite{Mantysaari:2017dwh,Sambasivam:2019gdd}. 

In this work, we go beyond previous studies \cite{Mantysaari:2016ykx,Mantysaari:2016jaz}, where model parameters were constrained ``by eye'', and perform a Bayesian analysis to 
extract in a statistically rigorous manner the  non-perturbative parameter values allowed by the HERA data, and construct initial conditions for hadronic collisions that are compatible with the experimental DIS data.

This paper is organized as follows. In Section \ref{sec:vmprod} we review the calculation of coherent and incoherent exclusive vector meson production in the dipole picture and discuss the various aspects of our model for the proton target. In Section \ref{sec:bayesian} we explain the procedure for our Bayesian analysis. We present results in Section \ref{sec:results} and conclude in Section \ref{sec:conclusions}.

\section{Vector meson production at high energy}
\label{sec:vmprod}

In this work we calculate vector meson production in a framework similar to the one used in Refs.~\cite{Mantysaari:2016jaz,Mantysaari:2016ykx} (see also Refs.~\cite{Kumar:2021zbn,Cepila:2018zky,Traini:2018hxd,Cepila:2017nef,Cepila:2016uku}), and for completeness briefly review the calculation in this section.

At high energies, DIS processes can be conveniently described in the dipole picture in the rest frame of the target proton, and interaction with the target color field is described in the Color Glass Condensate (CGC) framework~\cite{Kovchegov:2012mbw,Iancu:2003xm,Gelis:2010nm,Albacete:2014fwa}.
In the proton rest frame, the lifetime of a fluctuation of the incoming virtual photon into a quark-antiquark dipole is much longer than the characteristic timescale of the dipole-target interaction. Consequently, the scattering amplitude can be factorized into a convolution of photon and vector meson wave functions and the dipole-target interaction. The scattering amplitude for exclusive vector meson $V$ production can then be written as~\cite{Kowalski:2006hc,Hatta:2017cte}
\begin{multline}
\label{eq:jpsi_amp}
    \mathcal{A}^{\gamma^*+p \to V + p} = 2i\int \dd[2]{\rt} \dd[2]{\bt}  \frac{\dd{z}}{4\pi} e^{-i \left[\bt - \left(\frac{1}{2}-z\right)\rt\right]\cdot \Deltat} \\
    \times [\Psi_V^* \Psi_\gamma](Q^2,\rt,z) N_\Omega(\rt,\bt,\xpom).
\end{multline}
Here $\rt$ is the transverse size of the $q\bar q$ dipole,  $\bt$ is the impact parameter measured relative to the proton center, and $Q^2$ is the photon virtuality. The fraction of the large photon plus momentum carried by the quark is given by $z$, and $\Deltat$ is the transverse momentum transfer. Note that at high energies we can employ the eikonal approximation and assume that the quark transverse coordinates are fixed during the propagation through the target color field. 

The $\gamma^* \to q\bar q$ splitting is described by the virtual photon light front wave function $\Psi_\gamma$, which can be computed from QED~\cite{Kovchegov:2012mbw}. The vector meson wave function is non-perturbative, and in this work, we use the so-called Boosted Gaussian parametrization from~\cite{Kowalski:2006hc}, where the model parameters are constrained by experimental data on the vector meson decay width. We note that there are multiple vector meson wave functions proposed in the literature (see e.g.~Refs.~\cite{Lappi:2020ufv,Li:2017mlw,Li:2021cwv}). Different wave functions mostly affect the overall normalization of the \jpsi production cross section, and have a much smaller effect on the $|t|$ spectra, which we are most interested here~\cite{Lappi:2020ufv,Mantysaari:2017dwh,Kowalski:2006hc}. Consequently, our results will depend only weakly on the specific wave function choice (except for the parameter controlling the overall proton density).

Equation~\eqref{eq:jpsi_amp} is a leading order result for the vector meson production in the CGC framework (note that multiple scattering effects are resummed in the dipole amplitude $N_\Omega$). Currently, there is rapid progress in the field toward next-to-leading order (NLO) accuracy. In particular, the cross section for the production of  light mesons~\cite{Boussarie:2016bkq} and longitudinally polarized heavy vector mesons are now available~\cite{Mantysaari:2021ryb} at next-to-leading order, as well as the virtual photon light front wave function~\cite{Beuf:2021srj,Beuf:2020dxl,Beuf:2021qqa,Hanninen:2017ddy,Beuf:2017bpd,Beuf:2016wdz} and small-$x$ evolution equations~\cite{Lappi:2020srm,Lappi:2016fmu,Lappi:2015fma,Balitsky:2008zza,Ducloue:2019ezk,Ducloue:2019jmy,Iancu:2015vea,Iancu:2015joa,Balitsky:2013fea,Kovner:2013ona} (see also Ref.~\cite{Caucal:2021ent} where dijet production is studied at next-to-leading order). However, the NLO calculations are not yet at the level where they can be consistently used in phenomenological applications (in particular the cross section for transversely polarized heavy vector meson production is still missing). As the purpose of this work is to demonstrate the potential of Bayesian analyses to systematically extract non-perturbative parameters describing the proton event-by-event fluctuating structure, we do not expect the NLO contributions to have a large effect on the results we obtain using our leading order setup.

The coherent cross section, corresponding to the process where the target proton remains in the same quantum state, can be obtained by averaging over the target color charge configurations $\Omega$ at the amplitude level~\cite{Good:1960ba}:
\begin{equation}
     \frac{\dd \sigma^{\gamma^* + p \to V + p}}{\dd |t|}  = \frac{1}{16\pi} \left|\left\langle \mathcal{A}^{\gamma^*+p \to V + p} \right\rangle_\Omega\right|^2
\end{equation}

Subtracting the coherent contribution from the total diffractive vector meson production cross section we obtain the cross section for incoherent vector meson production, in which case the final state of the target is different from the initial state~\cite{Miettinen:1978jb,Caldwell:2010zza,Mantysaari:2020axf}. Experimentally, this corresponds to processes where the target proton (or nucleus) dissociates, but the rapidity gap between the produced vector meson and the target remnants remains. The incoherent cross section can then be written as a variance
\begin{multline}
     \frac{\dd \sigma^{\gamma^* + p \to V + p^*}}{\dd |t|}  = \frac{1}{16\pi} \left[
     \left\langle \left|\mathcal{A}^{\gamma^* + p \to V + p}\right|^2\right \rangle_\Omega \right. \\
     - \left. \left|\left\langle \mathcal{A}^{\gamma^*+p \to V + p} \right\rangle_\Omega\right|^2 \right]\,.
\end{multline}

Dependence on the small-$x$ structure of the target proton is included in the dipole amplitude $N_\Omega(\rt,\bt,\xpom)$, which, for a given target color charge configuration $\Omega$, can be written as 
\begin{equation}
    N_{\Omega}(\rt,\bt,\xpom) =  1 - \frac{1}{\nc} \tr \left[ V\left(\bt + \frac{\rt}{2}\right) V^\dagger\left(\bt - \frac{\rt}{2}\right) \right]. 
\end{equation}
Here $V(\xt)$ represents a Wilson line, which describes the color rotation of a quark state when it propagates through the target field (given the target color field configuration $\Omega$) at transverse coordinate $\xt$. We suppressed the dependence of $V$ on 
\begin{equation}
\label{eq:xpom}
    \xpom \approx \frac{M_V^2+Q^2}{W^2+Q^2}\,,
\end{equation}
which is the fraction of the target longitudinal momentum transferred to the meson with mass $M_V$ in the frame where the target has a large momentum. We neglect the $t$ dependence of $\xpom$ in this work. The photon-proton center-of-mass energy is denoted by $W$.

The Wilson lines are obtained in the same way as done in the IP-Glasma calculation~\cite{Schenke:2012wb} used e.g.~in Refs.~\cite{Mantysaari:2020lhf,Mantysaari:2019jhh,Mantysaari:2019csc,Mantysaari:2018zdd,Mantysaari:2016jaz,Mantysaari:2016ykx}. The color charges $\rho^a$ are first determined using the McLerran-Venugopalan~\cite{McLerran:1993ni} model, assuming that color charges are local Gaussian random variables with expectation value zero and variance
\begin{multline}
    g^2 \left \langle \rho^a(x^-, \xt) \rho^b(y^-,\yt) \right \rangle = g^4 \lambda_A(x^-) \delta^{ab} \\
    \times  \delta^{(2)}(\xt-\yt) \delta(x^- - y^-) \, .   
\end{multline}
Here the color charge density is $\mu^2 = \int \dd{x^-} \lambda_A(x^-)$, and it is related to the local saturation scale $Q_s(\xt)$ determined from the IPsat parametrization~\cite{Kowalski:2003hm} fitted to HERA data~\cite{Rezaeian:2012ji,Mantysaari:2018nng}. In our Bayesian analysis, the ratio
\begin{equation}
    \frac{Q_s(\xt)}{g^2\mu}\,,
\end{equation}
is a free parameter, controlling the overall proton density (see also Ref.~\cite{Lappi:2007ku} for a detailed study of this ratio). 
The Wilson lines $V(\xt)$ can be obtained by solving the Yang-Mills equations, and the result reads
\begin{equation}
  V(\xt) = \mathrm{P}_{-}\left\{ \exp\left({-ig\int_{-\infty}^\infty \dd{z^{-}} \frac{\rho^a(x^-,\xt) t^a}{\boldsymbol{\nabla}^2 - m^2} }\right) \right\}\,,
  \label{eq:wline_regulated}
\end{equation}
where $P_-$ represents path ordering in the $x^-$ direction. Here, we introduced the infrared regulator $m$, which is needed to avoid the emergence of unphysical Coulomb tails, and will be another free parameter in the Bayesian analysis.

In the IPsat parametrization the saturation scale $Q_s^2(\bt)$ is directly proportional to the local density $T_p(\bt)$. We introduce an event-by-event fluctuating density by writing the density profile following Refs.~\cite{Mantysaari:2016jaz,Mantysaari:2016ykx} as:
\begin{equation}
\label{eq:Tpfluct}
    T_p(\bt) = \frac{1}{N_q} \sum_{i=1 }^{N_q} p_i T_q(\bt-\bti),
\end{equation}
where 
\begin{equation}
    T_q(\bt) = \frac{1}{2\pi B_q} e^{-{\mathbf b}_\perp^2/(2B_q)}\,,
\end{equation}
and the coefficient $p_i$ allows for different normalizations for individual hot spots, to be discussed below.
Our prescription corresponds to having $N_q$ hot spots with hot spot width $B_q$ (note that the  hot spot transverse root mean square (RMS) radius in $\sqrt{2B_q}$). The hot spot positions $\bti$ are sampled from a two-dimensional Gaussian distribution whose width is denoted by $B_{qc}$, and the center-of-mass is shifted to the origin in the end.

As discussed in Ref.~\cite{Albacete:2016pmp,Albacete:2017ajt}, repulsive short-range correlations between the hot spots may explain the hollowness effect~\cite{Alkin:2014rfa,Dremin:2015ujt,Troshin:2016frs,Arriola:2016bxa} and negative correlation between the $v_2$ and $v_3$ flow harmonics observed in highest multiplicity proton-proton collisions~\cite{Sirunyan:2017uyl,Acharya:2019vdf}. In order to study if exclusive vector meson production in DIS can be used to probe such  repulsive correlations, we also introduce an additional model parameter $d_{q,\text{Min}}$ which controls the minimum three-dimensional distance required between any two hot spots.\footnote{To implement the minimal distance we follow \cite{Moreland:2014oya}, first sampling 3D distributions and if necessary resampling the solid angle until the requirement posed by $d_{q,\text{Min}}$ is satisfied.}
We checked that for a large number of hot spots $N_q = 10$ in a typical nucleon of size $B_{qc} = 4.2$ GeV$^{-2}$, the model parameter $d_{q, \mathrm{Min}}$ remains effective more than 90\% for $d_{q, \mathrm{Min}} \leq 0.4$ fm, meaning that in 10\% of the sampled configurations the distance requirement cannot be fulfilled.

Finally, we include saturation scale fluctuations by allowing the local density of each hot spot to fluctuate independently, following again Refs.~\cite{Mantysaari:2016jaz,Mantysaari:2016ykx} (see also Ref.~\cite{McLerran:2015qxa}). These fluctuations are implemented by sampling the coefficients $p_i$ in Eq.~\eqref{eq:Tpfluct} from the log-normal distribution
\begin{equation}
\label{eq:qsfluct}
P\left( \ln p_i \right) = \frac{1}{\sqrt{2\pi}\sigma} \exp \left[- \frac{\ln^2 p_i}{2\sigma^2}\right]\,.
\end{equation}
The sampled $p_i$ are at the end normalized by the expectation value of the distribution  $E[p_i]=e^{\sigma^2/2}$ in order to keep the average density unmodified. The magnitude of density fluctuations is controlled by the parameter $\sigma$.

\section{Bayesian analysis setup}
\label{sec:bayesian}

Bayesian Inference is a general and systematic method to constrain the probability distribution of model parameters $\btheta$ by comparing model calculations $\by(\btheta)$ with experimental measurements $\by_\mathrm{exp}$~\cite{sivia2006data}. Bayes' theorem provides the posterior distribution of model parameters as
\begin{equation}
    \CalP(\btheta \vert \by_\mathrm{exp}) \propto \CalP(\by_\mathrm{exp} \vert \btheta) \CalP(\btheta).
\end{equation}
Here $\CalP(\by_\mathrm{exp} \vert \btheta)$ is the likelihood for model results with parameter $\btheta$ to agree with the experimental data. We choose a multivariate normal distribution for the logarithm of the likelihood with $\Delta \by(\btheta) = \by(\btheta) - \by_\mathrm{exp}$ \cite{williams2006gaussian},
\begin{eqnarray}
    \ln[\CalP(\by_\mathrm{exp} \vert \btheta)] &=& - \frac{1}{2} \Delta \by(\btheta)^T \Sigma^{-1} \Delta \by(\btheta) \nonumber \\ && -\frac{1}{2} \ln[(2\pi)^n \det \Sigma].
\end{eqnarray}
Here $n$ is the number of experimental data points and $\Sigma \equiv \Sigma_\mathrm{exp} + \Sigma_\mathrm{model}$ is the $n \times n$ covariance matrix, which encodes experimental and model uncertainties.
In the current analysis, we assume no correlation among experimental errors of the $n$ observables $\{\sigma_i\}$. So the covariance matrix for experimental uncertainty takes a diagonal form,
\begin{equation}
    \Sigma_\mathrm{exp} = \mathrm{diag}(\sigma_1^2, \cdots, \sigma_n^2).
\end{equation}
Our model uncertainties $\Sigma_\mathrm{model}$ are estimated using the covariance matrix from the trained Gaussian Process (GP) Emulators \cite{Bernhard:2015hxa}.

\begin{table*}[t]
  \centering
  \caption{Summary of model parameters, their prior ranges, and constrained  maximum likelihood values with uncertainty estimates in 90\% credible intervals.
  }
  \begin{tabular}{lllll}\hline \hline
    Parameter & Description & Prior range & MAP (variable $N_q$) & MAP ($N_q\equiv3$) \\
    \hline
    $m$ $[\gev]$ & Infrared regulator & [0.05, 2] &  $0.506^{+1.12}_{-0.356}$ & $0.246^{+0.162}_{-0.103}$ \\ 
    $B_{qc}$ $[\gev^{-2}]$ & Proton size & [1, 10] & $4.02 ^{+1.73}_{-0.728}$ & $4.45^{+0.801}_{-0.803}$\\
    $B_q$ $[\gev^{-2}]$ & Hot spot size & [0.1, 3] & $0.474 ^{+0.434}_{-0.286}$ & $0.346^{+0.282}_{-0.202}$ \\
    $\sigma$ & Magnitude of $Q_s$ fluctuations & [0, 1.5] & $0.833 ^{+0.194}_{-0.441}$ & $0.563^{+0.143}_{-0.141}$ \\
    $Q_s/(g^2\mu)$ & Ratio of color charge density and saturation scale & [0.2, 1.5] & $0.598 ^{+0.230}_{-0.264}$ & $0.747^{+0.0704}_{-0.0930}$\\
    $d_{q,\text{Min}}$ $[\mathrm{fm}]$ & Minimum 3D distance between hot spots & [0, 0.5] & $0.257 ^{+0.221}_{-0.231}$ & $0.254^{+0.222}_{-0.229}$ \\
    $N_q$ & Number of hot spots & [1, 10] & $6.79 ^{+2.93}_{-4.83}$ & $3$ \\ \hline
  \end{tabular}
  \label{tab:modelparams}
\end{table*}
We employ GP emulators \cite{williams2006gaussian} for our model and couple them with the Monte-Carlo Markov Chain (MCMC) method to efficiently explore the model parameter space~\cite{goodman2010ensemble, foreman2013emcee}. The HERA measurements can be represented by five Principal Components (PC) with a residual variance of less than 0.01\%, meaning that 99.99\% of the variation of all studied observables within the prior parameter range are captured by the five principal components. Our GP emulators are trained to fit these five PCs with 1,000 training simulations in the model parameter space. In each model parameter point, we generate 3,000 configurations to compute the coherent and incoherent cross sections. The relative statistical errors are within 5\%.

All model parameters and their prior ranges (and Maximum a Posterior values that are discussed in Sec.~\ref{sec:results}) are summarized in Table.~\ref{tab:modelparams}.
We treat the parameter $N_q$ as a continuous real number. The fractional part of $N_q$ is treated as a probability to sample either $\lceil N_q \rceil$ or $\lfloor N_q \rfloor$ partons inside protons. The same approach was recently used in Ref.~\cite{Nijs:2021clz}.
The experimental data included in the Bayesian analysis is the H1 data on coherent and incoherent \jpsi production cross section measured at $W=75\,\gev$~\cite{H1:2013okq}. The incoherent data is included in the $|t|$ range $0<|t|< 2.5$ $\gev^2$. We note that there is incoherent data at higher $|t|$ also (studied in a similar context in Ref.~\cite{Kumar:2021zbn}), but the highest $|t|$ points are not included in our analysis for two reasons. First, as we determine Wilson lines at fixed $\xpom$, as discussed in Sec.~\ref{sec:vmprod}, and do not include the full $\xpom$ evolution, we neglect $|t|$ dependence in $\xpom$ (see Eq.~\eqref{eq:xpom}). Additionally, at large $|t|$ other effects such as DGLAP evolution~\cite{Gribov:1972ri,Gribov:1972rt,Altarelli:1977zs,Dokshitzer:1977sg} may become important.

\section{Results}
\label{sec:results}

\begin{figure*}[htb]
    \centering
    \includegraphics[width=\textwidth]{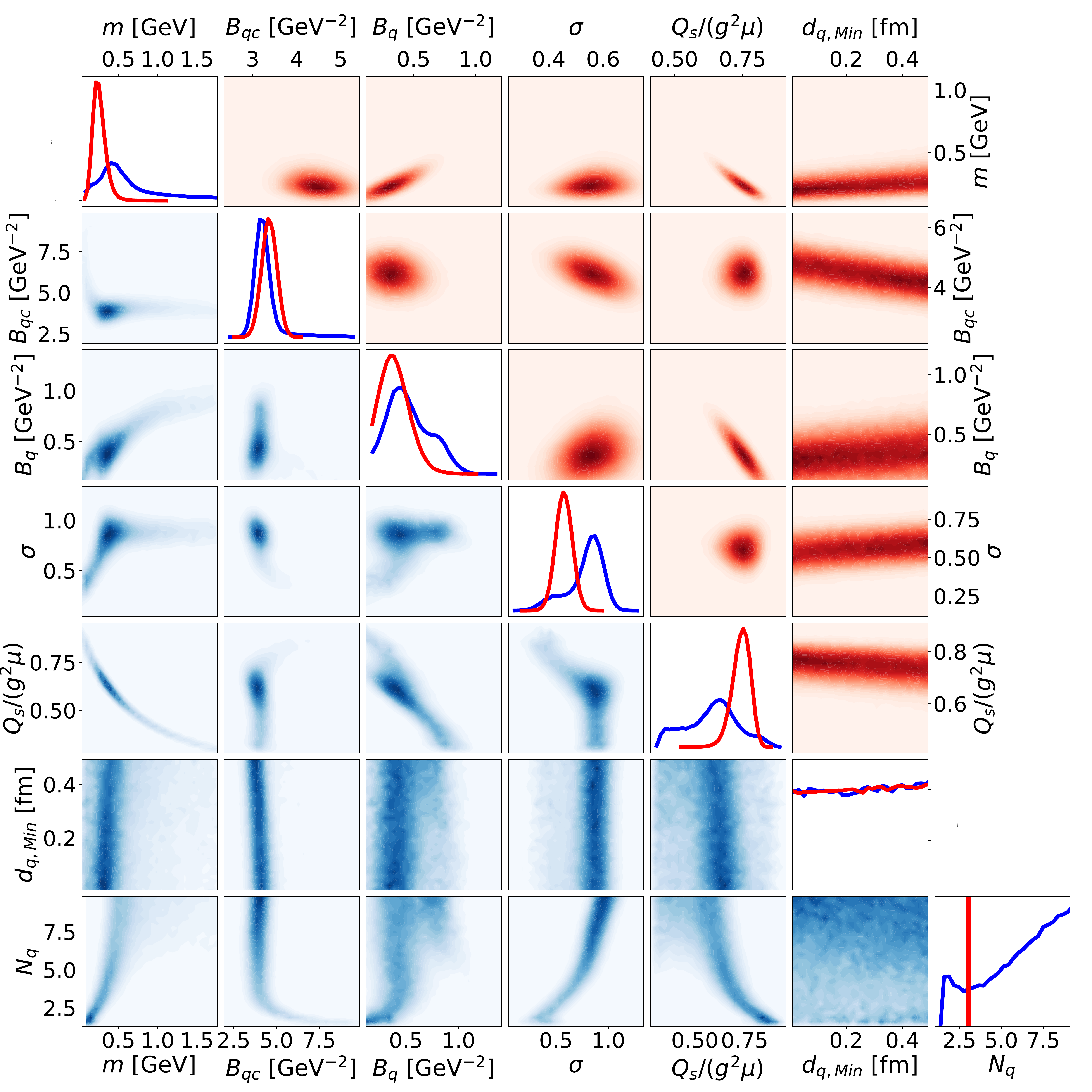}
    \caption{Bayesian posterior distributions of the model parameters. The diagonal panels show the probability distributions for individual parameters, and off-diagonal panels illustrate their pairwise correlations.  }
    \label{fig:posterior}
\end{figure*}

For two distinct scenarios, the first with $N_q$ fixed to 3, the second with $N_q$ a free parameter, the posterior distribution of model parameters is shown in Fig.~\ref{fig:posterior}.
Particularly for the case of $N_q=3$, most of the model parameters are tightly constrained by the H1 data included in the Bayesian analysis. This is due to the fact that different regions of the dataset are sensitive to different model parameters. 

First, the infrared regulator $m$ suppresses long-distance Coulomb tails, and as such, it controls the shape of the proton at large distances. This part of the proton geometry is probed by coherent diffraction at low $|t| \lesssim 0.2\gev^2$~\cite{Mantysaari:2016jaz}: impact parameter and momentum transfers are Fourier conjugates, and consequently the low-$|t|$ region is sensitive to large distances and vice versa. On the other hand, the actual proton size controlled mostly by $B_{qc}$ determines the overall slope of the coherent spectrum. The hot spot size $B_{q}$ then determines the slope of the incoherent cross section in the $|t|\gtrsim 1\,\gev^2$ region: as shown in Ref.~\cite{Lappi:2010dd} the slope of the incoherent spectra at high $|t|$ is given by the size of the smallest fluctuating constituent.

The overall normalization is determined by the $Q_s/(g^2\mu)$ parameter and the magnitude of the cross sections constrains that, however, it can not be determined very precisely from our analysis as it is strongly correlated with many other parameters, particularly for the case that $N_q$ is a free parameter. Here we note that there is some model uncertainty related to the non-perturbative vector meson wave functions, and different phenomenological parametrizations can result in cross sections that differ by $\sim 20\%$, see e.g.~\cite{Lappi:2020ufv,Mantysaari:2017dwh,Kowalski:2006hc}. 
In phenomenological analyses, the so-called  skewness correction~\cite{Kowalski:2006hc} is sometimes included, which can also have a numerically significant (up to $\sim 40\%$) effect on the cross section. However, as there are also other model uncertainties related to the overall normalization and the applicability of the skewness correction in our setup is not rigorously justified, it is not included in this work. The small real part correction discussed e.g. in Ref.~\cite{Kowalski:2006hc} is also neglected.

One can see that when leaving $N_q$ variable, its value can not be constrained in our analysis, except that $N_q\ge 2$ is required in order to get geometry fluctuations that are necessary to describe the incoherent HERA data. At first one would expect the configurations with large $N_q$ to be so smooth that event-by-event fluctuations would not be enough to result in a large enough incoherent cross section. However, we note that there is a strong positive correlation between $N_q$ and $\sigma$, which means that large $N_q$ goes along with large hot spot density fluctuations. In this situation one can not really interpret $N_q$ as the number of hot spots, but one has to consider effective hot spots, that are generated dynamically from the sum of the $N_q$ constituents that are each strongly fluctuating in magnitude. This effective hot spot number will generally be smaller than $N_q$. For a quantitative analysis a hot spot finding prescription, similar to jet clustering algorithms, would be needed.
Additional constraints for $\sigma$ originate from the incoherent cross section at small $|t|$, which is sensitive to the fluctuations at long-distance scales, i.e., overall density fluctuations~\cite{Mantysaari:2016jaz,Mantysaari:2016ykx}.

The minimum distance required between the hot spots in three dimensions, $d_{q,\text{Min}}$, can not be constrained at all in our analysis. This parameter is also only very weakly correlated with the other parameters, emphasizing its limited effect on the results. This means that the \jpsi production data allows, but does not require, repulsive short-range correlations used e.g.~in Refs.~\cite{Albacete:2016pmp,Albacete:2017ajt}. We note that for large $N_q$ it is not always possible to fulfill the minimum distance requirement, reducing the effective value of $d_{q,{\rm Min}}$ that is actually employed.

Let us next study correlations between the model parameters in more detail. First, we observe that there is a clear positive correlation between the infrared regulator $m$ and the hot spot size $B_q$. This can be understood, as increasing $m$ suppresses color fields at large distances and as such results in smaller hot spots. Interestingly, we find no clear  correlation between the overall proton size parameter $B_{qc}$ and the infrared regulator $m$.  In fact, the proton size $B_{qc}$ shows no clear correlation with any of the other model parameters.
The infrared regulator $m$ is strongly anti-correlated with $Q_s/(g^2\mu)$. This is again easy to understand: large $m$ results in a reduction of the normalization of the cross section, which has to be compensated by using a larger color charge density which requires smaller $Q_s/(g^2\mu)$. 

We also find a strong negative correlation between the number of hot spots $N_q$ and the proton density $Q_s/(g^2\mu)$. This correlation can be understood by considering the effect of these parameters on the incoherent cross section. When the number of hot spots with the same size increases, the incoherent slope is not significantly affected but the normalization goes down as there are smaller fluctuations in the scattering amplitude due to the smoother proton profile. This would result in the incoherent cross section being underestimated, which has to be compensated by smaller $Q_s/(g^2\mu)$. 

The somewhat surprising negative correlation between the hot spot size $B_q$ and the proton density $Q_s/(g^2\mu)$ can also be understood. First, we note that large hot spots (large $B_q$) require a larger infrared regulator $m$ which effectively makes the hot spots smaller (notice a positive correlation between $B_q$ and $m$). Then, to compensate for the effect of the larger infrared regulator on the magnitude of the cross section, a smaller $Q_s/(g^2\mu)$ is needed.

The posterior distribution of parameters $B_{qc}$ and $B_q$, that determine the proton size, which can be quantified e.g.~by the two dimensional RMS radius, $r_{\rm rms}=\sqrt{2(B_{\rm qc}+B_{q})}$, are shown in the second and third diagonal panels of Fig.~\ref{fig:posterior}. We find that they are sharply peaked, in particular for the case of fixed $N_q=3$. The extracted proton radius from this Bayesian analysis is $r_{\rm rms}=0.591^{+0.130}_{-0.071}\,{\rm fm}$ ($0.610^{+0.065}_{-0.068}\,{\rm fm}$) for variable $N_q$ ($N_q\equiv3$) with uncertainty estimates in 90\% credible intervals.  

We further included the possibility of fluctuating values of $B_q$ and $N_q$, and studied the dependence of the observables on the respective variances. We found that the considered experimental data could not constrain these parameters and decided not to include them in the presented analysis.

Next, we demonstrate explicitly that sampling parameter values from the posterior distribution lead to a good description of the HERA data. The main result of the Bayesian analysis is the posterior distribution shown in Fig.~\ref{fig:posterior}, but it is also possible to find the so-called Maximum a Posteriori (MAP) parameter set, which is the mode of the posterior distribution. Because our prior parameter distributions are uniform, the MAP parameters shown in Table~\ref{tab:modelparams} maximize the likelihood function and provide the best fit to the experimental data. We note that statistically the expectation value for any observable is not obtained using the MAP parameters. Instead, one should calculate observables using averages over parameter samples obtained from the posterior distribution.

\begin{figure}
    \centering
    \includegraphics[width=\columnwidth]{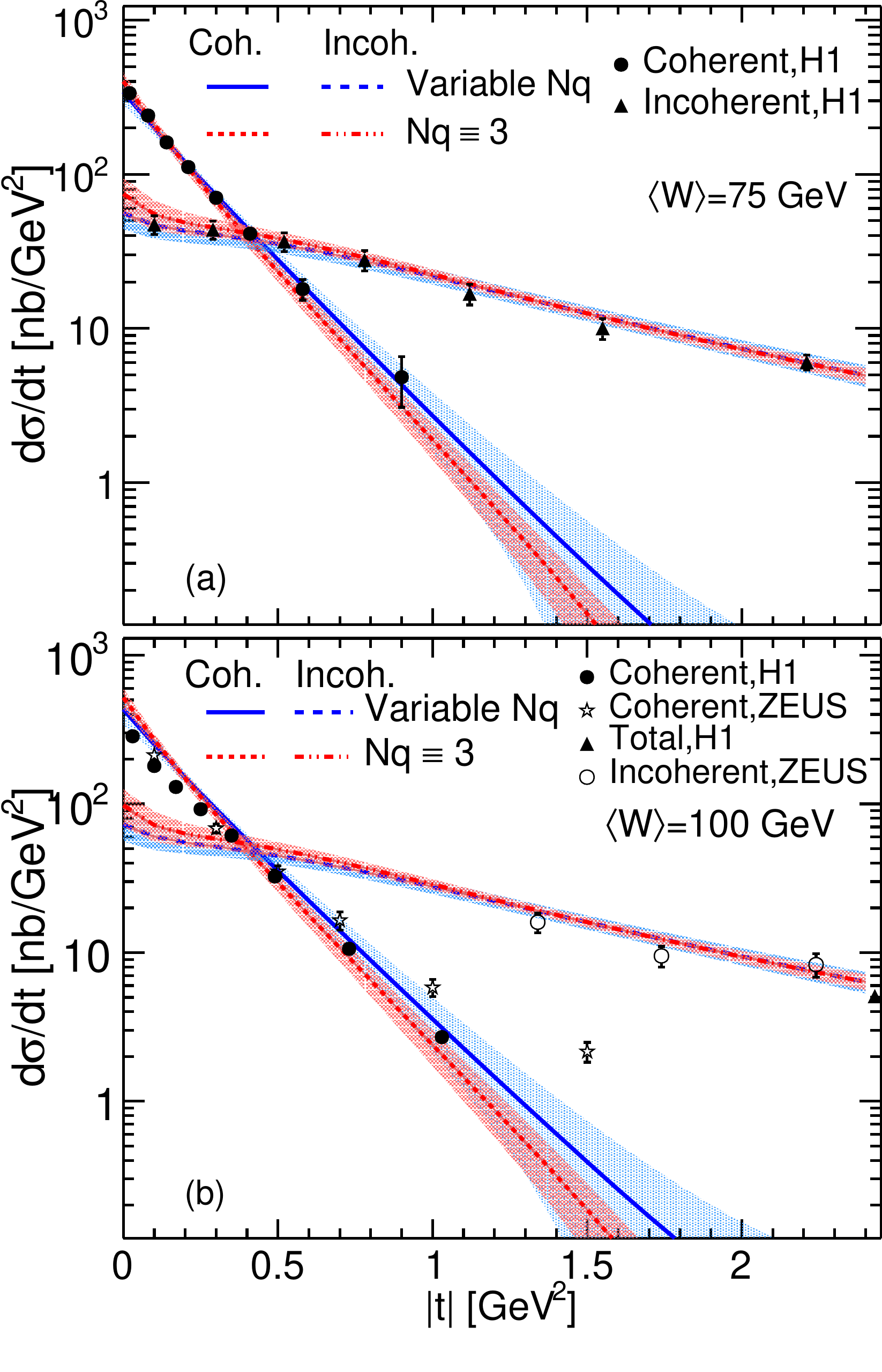}
    \caption{Coherent and incoherent \jpsi photoproduction cross sections at $\langle W \rangle = 75\,\gev$ (a) and $\langle W \rangle = 100\,\gev$ (b), calculated by averaging over many parameter sets sampled from the posterior distribution, which was determined using the  $\langle W \rangle = 75\,\gev$ HERA data~\cite{H1:2013okq}.  
    We compare the case with variable $N_q$ to that with fixed $N_q=3$, and to experimental data at  $\langle W \rangle = 75\,\gev$~\cite{H1:2013okq} and $\langle W \rangle = 100\,\gev$~\cite{Aktas:2005xu,Chekanov:2002xi,Chekanov:2002rm,Aktas:2003zi}, respectively. The bands show the one standard deviation uncertainty. }
    \label{fig:hera_w_75}
\end{figure}

Comparison to the HERA coherent and incoherent \jpsi production data measured at $W=75\,\gev$~\cite{H1:2013okq} and used in the Bayesian analysis to constrain the model parameters are shown in panel (a) of Fig.~\ref{fig:hera_w_75}. The spectra calculated by averaging the results computed using different parametrizations sampled from the posterior distribution indeed provide an excellent description of the data. We also show the statistical uncertainty, obtained by calculating the one standard deviation interval shown as red (coherent) and blue (incoherent) bands. 

Next, we study compatibility with experimental data not included in the Bayesian analysis.
We do not include full small-$x$ evolution e.g.~by means of the JIMWLK~\cite{Jalilian-Marian:1996mkd,Jalilian-Marian:1997qno, Jalilian-Marian:1997jhx,Iancu:2001md, Ferreiro:2001qy, Iancu:2001ad, Iancu:2000hn,Mueller:2001uk} equation in this work. Consequently, the only center-of-mass energy $W$ (or momentum fraction $\xpom$) dependence comes from the $\xpom$ dependence of the saturation scale $Q_s$ determined from the IPsat fit. This almost only affects the overall normalization of the calculated spectra, and misses important physical effects such as the growth of the proton with decreasing $x$. Consequently, we do not expect that within our setup we can describe \jpsi production at different center-of-mass energies simultaneously. Nevertheless, the geometry evolution between $W=75\,{\rm GeV}$ and $W=100\,{\rm GeV}$ should be weak enough to make predictions for the higher center-of-mass energy.

The calculated \jpsi spectra at $W=100\,{\rm GeV}$ are shown in panel (b) of Fig.~\ref{fig:hera_w_75} and compared with the data from H1 and ZEUS collaborations~\cite{Aktas:2005xu,ZEUS:2002wfj,Chekanov:2002rm,Aktas:2003zi}. Again, we show the model prediction as the average over many samples of the  posterior distribution and provide one standard deviation bands. The results are very similar to the $W=75\,\gev$ case studied above, and the HERA data is well described (except the coherent cross section at low $|t|$, and high $|t|$ in the case of the ZEUS measurement), and variation between the different parametrizations sampled from the posterior distribution is small. Similarity to the $W=75\,\gev$ case is not surprising, as we use exactly the same fluctuating geometry, with the only difference being slightly larger $Q_s$ values, extracted from the IPsat parametrization.

\section{Conclusions}
\label{sec:conclusions}

We have performed a statistically rigorous Bayesian analysis to extract posterior likelihood distributions for the non-perturbative parameters describing the event-by-event fluctuating proton geometry as constrained by exclusive \jpsi production data from HERA. We presented a comparison of the $|t|$-dependent coherent and incoherent cross sections, obtained from an average over many parameter samples from the posterior distribution, to the experimental data.

Generally, model parametrizations sampled from the determined posterior distribution can be used to systematically take into account uncertainties in the proton geometry as constrained by the HERA DIS data when calculating any other observable that depends on the proton geometry, such as flow observables in high-multiplicity proton-proton and proton-nucleus collisions. To enable such studies, we provide 1,000 parametrizations sampled from the determined posterior distributions in the supplemental material.  

Because the model parameters are sensitive to different aspects of the coherent and incoherent vector meson production spectra, most of them are well constrained by the Bayesian analysis. The only exceptions are the minimum distance between the hot spots (repulsive short-range correlations), and the number of hot spots $N_q$, which cannot be well constrained by the considered HERA data. However, we note that although the analysis suggests that large $N_q$ is compatible with the HERA data, for $N_q \gtrsim 5$ one should no longer simply interpret $N_q$ as the number of actual hot spots. In this regime, a good description of the data requires large fluctuations of individual hot spots' densities, and the sampled `hot spots' can also overlap significantly. Thus, the effective number of hot spots is significantly smaller than the parameter $N_q$ might imply.

The HERA data used in this work probes the proton structure at $x \sim 10^{-3}$. The energy (Bjorken-$x$) dependence can be included in terms of JIMWLK evolution as e.g.~in Refs.~\cite{Schlichting:2014ipa,Mantysaari:2018zdd}. In the future we plan to extend our framework by including the full JIMWLK evolution and vector meson production data at different center-of-mass energies from HERA~\cite{ZEUS:2002wfj,H1:2013okq,H1:2005dtp,H1:2013okq} and from the ultra peripheral proton-lead collisions measured at the LHC~\cite{Aaij:2014iea,Acharya:2018jua,LHCb:2014acg, LHCb:2018rcm}, allowing us to extract also the Bjorken-$x$ dependence of the fluctuating proton geometry. 

In addition to constraining the energy dependence, performing global analyses including both exclusive vector meson production data from HERA and flow harmonics from the proton-proton and proton-lead collisions measured at the LHC (including a model calculation along the lines of \cite{Mantysaari:2017cni}) would allow for a powerful global analysis of the fluctuating nucleon substructure and properties of the final state. More differential DIS measurements from the future EIC such as dijet~\cite{Mantysaari:2019csc} or lepton-meson angular correlations~\cite{Mantysaari:2020lhf} can also provide further constraints and can in principle be included in our framework in a straightforward manner.

The numerical framework for our physics models and the Bayesian analysis package are publicly available on Github~\cite{IPGlasmaDiffraction, BayesianPackage}. Our Bayesian analysis code is developed based on the open-source numerical package by the Duke group~\cite{Bernhard:2019bmu}. To help visualize how observables depend on the model parameters, we provide an interactive web page with the trained GP emulators \cite{StreamlitApp}, where one can also find the posterior samples.

\section*{Acknowledgments}
B.P.S. and C.S. are supported by the U.S. Department of Energy, Office of Science, Office of Nuclear Physics, under DOE Contract No.~DE-SC0012704 and Award No.~DE-SC0021969, respectively.  C.S. acknowledges a DOE Office of Science Early Career Award.
H.M. is supported by the Academy of Finland, the Centre of Excellence in Quark Matter, and projects 338263 and 346567, and by the EU Horizon 2020 research and innovation programme, STRONG-2020 project (Grant Agreement No. 824093).
W.B.Z. is supported by the National Science Foundation (NSF) under grant numbers ACI-2004571 within the framework of the XSCAPE project of the JETSCAPE collaboration.
The content of this article does not reflect the official opinion of the European Union and responsibility for the information and views expressed therein lies entirely with the authors.
This research was done using resources provided by the Open Science Grid (OSG)~\cite{Pordes:2007zzb, Sfiligoi:2009cct}, which is supported by the National Science Foundation award \#2030508.

\bibliographystyle{JHEP-2modlong.bst}
\bibliography{spires}

\providecommand{\href}[2]{#2}\begingroup\raggedright\begin{thebibliography}{100}

\bibitem{AbdulKhalek:2021gbh}
R.~Abdul~Khalek {\em et.~al.}, {\it {Science Requirements and Detector Concepts
  for the Electron-Ion Collider: EIC Yellow Report}},
  \href{http://arXiv.org/abs/2103.05419}{{\tt arXiv:2103.05419
  [physics.ins-det]}}.

\bibitem{Aschenauer:2017jsk}
E.~C. Aschenauer, S.~Fazio, J.~H. Lee, H.~Mäntysaari, B.~S. Page, B.~Schenke,
  T.~Ullrich, R.~Venugopalan and P.~Zurita, {\it {The electron\textendash{}ion
  collider: assessing the energy dependence of key measurements}},
  \href{http://dx.doi.org/10.1088/1361-6633/aaf216}{{\em Rept. Prog. Phys.}
  {\bf 82} (2019)~no.~2 024301} [\href{http://arXiv.org/abs/1708.01527}{{\tt
  arXiv:1708.01527 [nucl-ex]}}].

\bibitem{Agostini:2020fmq}
{\bf LHeC, FCC-he Study Group} collaboration, P.~Agostini {\em et.~al.}, {\it
  {The Large Hadron-Electron Collider at the HL-LHC}},
  \href{http://dx.doi.org/10.1088/1361-6471/abf3ba}{{\em J. Phys. G} {\bf 48}
  (2021)~no.~11 110501} [\href{http://arXiv.org/abs/2007.14491}{{\tt
  arXiv:2007.14491 [hep-ex]}}].

\bibitem{Anderle:2021wcy}
D.~P. Anderle {\em et.~al.}, {\it {Electron-ion collider in China}},
  \href{http://dx.doi.org/10.1007/s11467-021-1062-0}{{\em Front. Phys.
  (Beijing)} {\bf 16} (2021)~no.~6 64701}
  [\href{http://arXiv.org/abs/2102.09222}{{\tt arXiv:2102.09222 [nucl-ex]}}].

\bibitem{Ryskin:1992ui}
M.~G. Ryskin, {\it {Diffractive $\mathrm{J}/\psi$ electroproduction in LLA
  QCD}},  \href{http://dx.doi.org/10.1007/BF01555742}{{\em Z. Phys. C} {\bf 57}
  (1993) 89}.

\bibitem{Mantysaari:2020axf}
H.~M\"antysaari, {\it {Review of proton and nuclear shape fluctuations at high
  energy}},  \href{http://dx.doi.org/10.1088/1361-6633/aba347}{{\em Rept. Prog.
  Phys.} {\bf 83} (2020)~no.~8 082201}
  [\href{http://arXiv.org/abs/2001.10705}{{\tt arXiv:2001.10705 [hep-ph]}}].

\bibitem{ALICE:2012eyl}
{\bf ALICE} collaboration, B.~Abelev {\em et.~al.}, {\it {Long-range angular
  correlations on the near and away side in $p$-Pb collisions at
  $\sqrt{s_{NN}}=5.02$ TeV}},
  \href{http://dx.doi.org/10.1016/j.physletb.2013.01.012}{{\em Phys. Lett. B}
  {\bf 719} (2013) 29} [\href{http://arXiv.org/abs/1212.2001}{{\tt
  arXiv:1212.2001 [nucl-ex]}}].

\bibitem{CMS:2012qk}
{\bf CMS} collaboration, S.~Chatrchyan {\em et.~al.}, {\it {Observation of
  Long-Range Near-Side Angular Correlations in Proton-Lead Collisions at the
  LHC}},  \href{http://dx.doi.org/10.1016/j.physletb.2012.11.025}{{\em Phys.
  Lett. B} {\bf 718} (2013) 795} [\href{http://arXiv.org/abs/1210.5482}{{\tt
  arXiv:1210.5482 [nucl-ex]}}].

\bibitem{ATLAS:2012cix}
{\bf ATLAS} collaboration, G.~Aad {\em et.~al.}, {\it {Observation of
  Associated Near-Side and Away-Side Long-Range Correlations in
  $\sqrt{s_{NN}}$=5.02 TeV Proton-Lead Collisions with the ATLAS Detector}},
  \href{http://dx.doi.org/10.1103/PhysRevLett.110.182302}{{\em Phys. Rev.
  Lett.} {\bf 110} (2013)~no.~18 182302}
  [\href{http://arXiv.org/abs/1212.5198}{{\tt arXiv:1212.5198 [hep-ex]}}].

\bibitem{PHENIX:2018lia}
{\bf PHENIX} collaboration, C.~Aidala {\em et.~al.}, {\it {Creation of
  quark\textendash{}gluon plasma droplets with three distinct geometries}},
  \href{http://dx.doi.org/10.1038/s41567-018-0360-0}{{\em Nature Phys.} {\bf
  15} (2019)~no.~3 214} [\href{http://arXiv.org/abs/1805.02973}{{\tt
  arXiv:1805.02973 [nucl-ex]}}].

\bibitem{STAR:2019zaf}
{\bf STAR} collaboration, J.~Adam {\em et.~al.}, {\it {Azimuthal Harmonics in
  Small and Large Collision Systems at RHIC Top Energies}},
  \href{http://dx.doi.org/10.1103/PhysRevLett.122.172301}{{\em Phys. Rev.
  Lett.} {\bf 122} (2019)~no.~17 172301}
  [\href{http://arXiv.org/abs/1901.08155}{{\tt arXiv:1901.08155 [nucl-ex]}}].

\bibitem{CMS:2010ifv}
{\bf CMS} collaboration, V.~Khachatryan {\em et.~al.}, {\it {Observation of
  Long-Range Near-Side Angular Correlations in Proton-Proton Collisions at the
  LHC}},  \href{http://dx.doi.org/10.1007/JHEP09(2010)091}{{\em JHEP} {\bf 09}
  (2010) 091} [\href{http://arXiv.org/abs/1009.4122}{{\tt arXiv:1009.4122
  [hep-ex]}}].

\bibitem{ATLAS:2021jhn}
{\bf ATLAS} collaboration, G.~Aad {\em et.~al.}, {\it {Two-particle azimuthal
  correlations in photonuclear ultraperipheral Pb+Pb collisions at 5.02 TeV
  with ATLAS}},  \href{http://dx.doi.org/10.1103/PhysRevC.104.014903}{{\em
  Phys. Rev. C} {\bf 104} (2021)~no.~1 014903}
  [\href{http://arXiv.org/abs/2101.10771}{{\tt arXiv:2101.10771 [nucl-ex]}}].

\bibitem{Dusling:2015gta}
K.~Dusling, W.~Li and B.~Schenke, {\it {Novel collective phenomena in
  high-energy proton\textendash{}proton and proton\textendash{}nucleus
  collisions}},  \href{http://dx.doi.org/10.1142/S0218301316300022}{{\em Int.
  J. Mod. Phys. E} {\bf 25} (2016)~no.~01 1630002}
  [\href{http://arXiv.org/abs/1509.07939}{{\tt arXiv:1509.07939 [nucl-ex]}}].

\bibitem{Loizides:2016tew}
C.~Loizides, {\it {Experimental overview on small collision systems at the
  LHC}},  \href{http://dx.doi.org/10.1016/j.nuclphysa.2016.04.022}{{\em Nucl.
  Phys. A} {\bf 956} (2016) 200} [\href{http://arXiv.org/abs/1602.09138}{{\tt
  arXiv:1602.09138 [nucl-ex]}}].

\bibitem{Schlichting:2016sqo}
S.~Schlichting and P.~Tribedy, {\it {Collectivity in Small Collision Systems:
  An Initial-State Perspective}},
  \href{http://dx.doi.org/10.1155/2016/8460349}{{\em Adv. High Energy Phys.}
  {\bf 2016} (2016) 8460349} [\href{http://arXiv.org/abs/1611.00329}{{\tt
  arXiv:1611.00329 [hep-ph]}}].

\bibitem{Nagle:2018nvi}
J.~L. Nagle and W.~A. Zajc, {\it {Small System Collectivity in Relativistic
  Hadronic and Nuclear Collisions}},
  \href{http://dx.doi.org/10.1146/annurev-nucl-101916-123209}{{\em Ann. Rev.
  Nucl. Part. Sci.} {\bf 68} (2018) 211}
  [\href{http://arXiv.org/abs/1801.03477}{{\tt arXiv:1801.03477 [nucl-ex]}}].

\bibitem{Schenke:2019pmk}
B.~Schenke, C.~Shen and P.~Tribedy, {\it {Hybrid Color Glass Condensate and
  hydrodynamic description of the Relativistic Heavy Ion Collider small system
  scan}},  \href{http://dx.doi.org/10.1016/j.physletb.2020.135322}{{\em Phys.
  Lett. B} {\bf 803} (2020) 135322}
  [\href{http://arXiv.org/abs/1908.06212}{{\tt arXiv:1908.06212 [nucl-th]}}].

\bibitem{Schenke:2021mxx}
B.~Schenke, {\it {The smallest fluid on Earth}},
  \href{http://dx.doi.org/10.1088/1361-6633/ac14c9}{{\em Rept. Prog. Phys.}
  {\bf 84} (2021)~no.~8 082301} [\href{http://arXiv.org/abs/2102.11189}{{\tt
  arXiv:2102.11189 [nucl-th]}}].

\bibitem{JETSCAPE:2021ehl}
{\bf JETSCAPE} collaboration, S.~Cao {\em et.~al.}, {\it {Determining the jet
  transport coefficient $\hat{q}$ from inclusive hadron suppression
  measurements using Bayesian parameter estimation}},
  \href{http://dx.doi.org/10.1103/PhysRevC.104.024905}{{\em Phys. Rev. C} {\bf
  104} (2021)~no.~2 024905} [\href{http://arXiv.org/abs/2102.11337}{{\tt
  arXiv:2102.11337 [nucl-th]}}].

\bibitem{JETSCAPE:2020mzn}
{\bf JETSCAPE} collaboration, D.~Everett {\em et.~al.}, {\it {Multisystem
  Bayesian constraints on the transport coefficients of QCD matter}},
  \href{http://dx.doi.org/10.1103/PhysRevC.103.054904}{{\em Phys. Rev. C} {\bf
  103} (2021)~no.~5 054904} [\href{http://arXiv.org/abs/2011.01430}{{\tt
  arXiv:2011.01430 [hep-ph]}}].

\bibitem{JETSCAPE:2020shq}
{\bf JETSCAPE} collaboration, D.~Everett {\em et.~al.}, {\it {Phenomenological
  constraints on the transport properties of QCD matter with data-driven model
  averaging}},  \href{http://dx.doi.org/10.1103/PhysRevLett.126.242301}{{\em
  Phys. Rev. Lett.} {\bf 126} (2021)~no.~24 242301}
  [\href{http://arXiv.org/abs/2010.03928}{{\tt arXiv:2010.03928 [hep-ph]}}].

\bibitem{Bernhard:2019bmu}
J.~E. Bernhard, J.~S. Moreland and S.~A. Bass, {\it {Bayesian estimation of the
  specific shear and bulk viscosity of quark\textendash{}gluon plasma}},
  \href{http://dx.doi.org/10.1038/s41567-019-0611-8}{{\em Nature Phys.} {\bf
  15} (2019)~no.~11 1113}.

\bibitem{Bernhard:2016tnd}
J.~E. Bernhard, J.~S. Moreland, S.~A. Bass, J.~Liu and U.~Heinz, {\it {Applying
  Bayesian parameter estimation to relativistic heavy-ion collisions:
  simultaneous characterization of the initial state and quark-gluon plasma
  medium}},  \href{http://dx.doi.org/10.1103/PhysRevC.94.024907}{{\em Phys.
  Rev. C} {\bf 94} (2016)~no.~2 024907}
  [\href{http://arXiv.org/abs/1605.03954}{{\tt arXiv:1605.03954 [nucl-th]}}].

\bibitem{Parkkila:2021yha}
J.~E. Parkkila, A.~Onnerstad, F.~Taghavi, C.~Mordasini, A.~Bilandzic and D.~J.
  Kim, {\it {New constraints for QCD matter from improved Bayesian parameter
  estimation in heavy-ion collisions at LHC}},
  \href{http://arXiv.org/abs/2111.08145}{{\tt arXiv:2111.08145 [hep-ph]}}.

\bibitem{Parkkila:2021tqq}
J.~E. Parkkila, A.~Onnerstad and D.~J. Kim, {\it {Bayesian estimation of the
  specific shear and bulk viscosity of the quark-gluon plasma with additional
  flow harmonic observables}},
  \href{http://dx.doi.org/10.1103/PhysRevC.104.054904}{{\em Phys. Rev. C} {\bf
  104} (2021)~no.~5 054904} [\href{http://arXiv.org/abs/2106.05019}{{\tt
  arXiv:2106.05019 [hep-ph]}}].

\bibitem{Moreland:2018gsh}
J.~S. Moreland, J.~E. Bernhard and S.~A. Bass, {\it {Bayesian calibration of a
  hybrid nuclear collision model using p-Pb and Pb-Pb data at energies
  available at the CERN Large Hadron Collider}},
  \href{http://dx.doi.org/10.1103/PhysRevC.101.024911}{{\em Phys. Rev. C} {\bf
  101} (2020)~no.~2 024911} [\href{http://arXiv.org/abs/1808.02106}{{\tt
  arXiv:1808.02106 [nucl-th]}}].

\bibitem{H1:2013okq}
{\bf H1} collaboration, C.~Alexa {\em et.~al.}, {\it {Elastic and
  Proton-Dissociative Photoproduction of $\mathrm{J}/\psi$ Mesons at HERA}},
  \href{http://dx.doi.org/10.1140/epjc/s10052-013-2466-y}{{\em Eur. Phys. J. C}
  {\bf 73} (2013)~no.~6 2466} [\href{http://arXiv.org/abs/1304.5162}{{\tt
  arXiv:1304.5162 [hep-ex]}}].

\bibitem{Aktas:2005xu}
{\bf H1} collaboration, A.~Aktas {\em et.~al.}, {\it {Elastic $\mathrm{J}/\psi$
  production at HERA}},
  \href{http://dx.doi.org/10.1140/epjc/s2006-02519-5}{{\em Eur. Phys. J. C}
  {\bf 46} (2006) 585} [\href{http://arXiv.org/abs/hep-ex/0510016}{{\tt
  arXiv:hep-ex/0510016}}].

\bibitem{Chekanov:2002xi}
{\bf ZEUS} collaboration, S.~Chekanov {\em et.~al.}, {\it {Exclusive
  photoproduction of $\mathrm{J}/\psi$ mesons at HERA}},
  \href{http://dx.doi.org/10.1007/s10052-002-0953-7}{{\em Eur. Phys. J. C} {\bf
  24} (2002) 345} [\href{http://arXiv.org/abs/hep-ex/0201043}{{\tt
  arXiv:hep-ex/0201043}}].

\bibitem{Chekanov:2002rm}
{\bf ZEUS} collaboration, S.~Chekanov {\em et.~al.}, {\it {Measurement of
  proton dissociative diffractive photoproduction of vector mesons at large
  momentum transfer at HERA}},
  \href{http://dx.doi.org/10.1140/epjc/s2002-01079-0}{{\em Eur. Phys. J. C}
  {\bf 26} (2003) 389} [\href{http://arXiv.org/abs/hep-ex/0205081}{{\tt
  arXiv:hep-ex/0205081}}].

\bibitem{Aktas:2003zi}
{\bf H1} collaboration, A.~Aktas {\em et.~al.}, {\it {Diffractive
  photoproduction of $\mathrm{J}/\psi$ mesons with large momentum transfer at
  HERA}},  \href{http://dx.doi.org/10.1016/j.physletb.2003.06.056}{{\em Phys.
  Lett. B} {\bf 568} (2003) 205}
  [\href{http://arXiv.org/abs/hep-ex/0306013}{{\tt arXiv:hep-ex/0306013}}].

\bibitem{Mantysaari:2016ykx}
H.~M\"antysaari and B.~Schenke, {\it {Evidence of strong proton shape
  fluctuations from incoherent diffraction}},
  \href{http://dx.doi.org/10.1103/PhysRevLett.117.052301}{{\em Phys. Rev.
  Lett.} {\bf 117} (2016)~no.~5 052301}
  [\href{http://arXiv.org/abs/1603.04349}{{\tt arXiv:1603.04349 [hep-ph]}}].

\bibitem{Klein:2019qfb}
S.~R. Klein and H.~M\"antysaari, {\it {Imaging the nucleus with high-energy
  photons}},  \href{http://dx.doi.org/10.1038/s42254-019-0107-6}{{\em Nature
  Rev. Phys.} {\bf 1} (2019)~no.~11 662}
  [\href{http://arXiv.org/abs/1910.10858}{{\tt arXiv:1910.10858 [hep-ex]}}].

\bibitem{Bertulani:2005ru}
C.~A. Bertulani, S.~R. Klein and J.~Nystrand, {\it {Physics of ultra-peripheral
  nuclear collisions}},
  \href{http://dx.doi.org/10.1146/annurev.nucl.55.090704.151526}{{\em Ann. Rev.
  Nucl. Part. Sci.} {\bf 55} (2005) 271}
  [\href{http://arXiv.org/abs/nucl-ex/0502005}{{\tt arXiv:nucl-ex/0502005}}].

\bibitem{PHENIX:2009xtn}
{\bf PHENIX} collaboration, S.~Afanasiev {\em et.~al.}, {\it {Photoproduction
  of $\mathrm{J}/\psi$ and of high mass $e^+e^-$ in ultra-peripheral Au+Au
  collisions at $\sqrt{s}= 200$ GeV}},
  \href{http://dx.doi.org/10.1016/j.physletb.2009.07.061}{{\em Phys. Lett. B}
  {\bf 679} (2009) 321} [\href{http://arXiv.org/abs/0903.2041}{{\tt
  arXiv:0903.2041 [nucl-ex]}}].

\bibitem{ALICE:2014eof}
{\bf ALICE} collaboration, B.~B. Abelev {\em et.~al.}, {\it {Exclusive
  $\mathrm{J/}\psi$ photoproduction off protons in ultra-peripheral p-Pb
  collisions at $\sqrt{s_{\rm NN}}=5.02$ TeV}},
  \href{http://dx.doi.org/10.1103/PhysRevLett.113.232504}{{\em Phys. Rev.
  Lett.} {\bf 113} (2014)~no.~23 232504}
  [\href{http://arXiv.org/abs/1406.7819}{{\tt arXiv:1406.7819 [nucl-ex]}}].

\bibitem{ALICE:2018oyo}
{\bf ALICE} collaboration, S.~Acharya {\em et.~al.}, {\it {Energy dependence of
  exclusive $\mathrm {J}/\psi $ photoproduction off protons in ultra-peripheral
  p\textendash{}Pb collisions at $\sqrt{s_{\mathrm {\scriptscriptstyle NN}}} =
  5.02$ TeV}},  \href{http://dx.doi.org/10.1140/epjc/s10052-019-6816-2}{{\em
  Eur. Phys. J. C} {\bf 79} (2019)~no.~5 402}
  [\href{http://arXiv.org/abs/1809.03235}{{\tt arXiv:1809.03235 [nucl-ex]}}].

\bibitem{LHCb:2014acg}
{\bf LHCb} collaboration, R.~Aaij {\em et.~al.}, {\it {Updated measurements of
  exclusive $J/\psi$ and $\psi$(2S) production cross-sections in pp collisions
  at $\sqrt{s}=7$ TeV}},
  \href{http://dx.doi.org/10.1088/0954-3899/41/5/055002}{{\em J. Phys. G} {\bf
  41} (2014) 055002} [\href{http://arXiv.org/abs/1401.3288}{{\tt
  arXiv:1401.3288 [hep-ex]}}].

\bibitem{LHCb:2018rcm}
{\bf LHCb} collaboration, R.~Aaij {\em et.~al.}, {\it {Central exclusive
  production of $J/\psi$ and $\psi(2S)$ mesons in $pp$ collisions at
  $\sqrt{s}=13~$TeV}},  \href{http://dx.doi.org/10.1007/JHEP10(2018)167}{{\em
  JHEP} {\bf 10} (2018) 167} [\href{http://arXiv.org/abs/1806.04079}{{\tt
  arXiv:1806.04079 [hep-ex]}}].

\bibitem{CMS:2016itn}
{\bf CMS} collaboration, V.~Khachatryan {\em et.~al.}, {\it {Coherent $J/\psi$
  photoproduction in ultra-peripheral PbPb collisions at $\sqrt {s_{NN}} =$
  2.76 TeV with the CMS experiment}},
  \href{http://dx.doi.org/10.1016/j.physletb.2017.07.001}{{\em Phys. Lett. B}
  {\bf 772} (2017) 489} [\href{http://arXiv.org/abs/1605.06966}{{\tt
  arXiv:1605.06966 [nucl-ex]}}].

\bibitem{ALICE:2021tyx}
{\bf ALICE} collaboration, S.~Acharya {\em et.~al.}, {\it {First measurement of
  the $|t|$-dependence of coherent $\mathrm{J}/\psi$ photonuclear production}},
   \href{http://dx.doi.org/10.1016/j.physletb.2021.136280}{{\em Phys. Lett. B}
  {\bf 817} (2021) 136280} [\href{http://arXiv.org/abs/2101.04623}{{\tt
  arXiv:2101.04623 [nucl-ex]}}].

\bibitem{Mantysaari:2017dwh}
H.~M\"antysaari and B.~Schenke, {\it {Probing subnucleon scale fluctuations in
  ultraperipheral heavy ion collisions}},
  \href{http://dx.doi.org/10.1016/j.physletb.2017.07.063}{{\em Phys. Lett. B}
  {\bf 772} (2017) 832} [\href{http://arXiv.org/abs/1703.09256}{{\tt
  arXiv:1703.09256 [hep-ph]}}].

\bibitem{Sambasivam:2019gdd}
B.~Sambasivam, T.~Toll and T.~Ullrich, {\it {Investigating saturation effects
  in ultraperipheral collisions at the LHC with the color dipole model}},
  \href{http://dx.doi.org/10.1016/j.physletb.2020.135277}{{\em Phys. Lett. B}
  {\bf 803} (2020) 135277} [\href{http://arXiv.org/abs/1910.02899}{{\tt
  arXiv:1910.02899 [hep-ph]}}].

\bibitem{Mantysaari:2016jaz}
H.~M\"antysaari and B.~Schenke, {\it {Revealing proton shape fluctuations with
  incoherent diffraction at high energy}},
  \href{http://dx.doi.org/10.1103/PhysRevD.94.034042}{{\em Phys. Rev. D} {\bf
  94} (2016)~no.~3 034042} [\href{http://arXiv.org/abs/1607.01711}{{\tt
  arXiv:1607.01711 [hep-ph]}}].

\bibitem{Kumar:2021zbn}
A.~Kumar and T.~Toll, {\it {Investigating the structure of gluon fluctuations
  in the proton with incoherent diffraction at HERA}},
  \href{http://arXiv.org/abs/2106.12855}{{\tt arXiv:2106.12855 [hep-ph]}}.

\bibitem{Cepila:2018zky}
J.~Cepila, J.~G. Contreras, M.~Krelina and J.~D. Tapia~Takaki, {\it {Mass
  dependence of vector meson photoproduction off protons and nuclei within the
  energy-dependent hot-spot model}},
  \href{http://dx.doi.org/10.1016/j.nuclphysb.2018.07.010}{{\em Nucl. Phys. B}
  {\bf 934} (2018) 330} [\href{http://arXiv.org/abs/1804.05508}{{\tt
  arXiv:1804.05508 [hep-ph]}}].

\bibitem{Traini:2018hxd}
M.~C. Traini and J.-P. Blaizot, {\it {Diffractive incoherent vector meson
  production off protons: a quark model approach to gluon fluctuation
  effects}},  \href{http://dx.doi.org/10.1140/epjc/s10052-019-6826-0}{{\em Eur.
  Phys. J. C} {\bf 79} (2019)~no.~4 327}
  [\href{http://arXiv.org/abs/1804.06110}{{\tt arXiv:1804.06110 [hep-ph]}}].

\bibitem{Cepila:2017nef}
J.~Cepila, J.~G. Contreras and M.~Krelina, {\it {Coherent and incoherent
  $\mathrm{J/}\psi$ photonuclear production in an energy-dependent hot-spot
  model}},  \href{http://dx.doi.org/10.1103/PhysRevC.97.024901}{{\em Phys. Rev.
  C} {\bf 97} (2018)~no.~2 024901} [\href{http://arXiv.org/abs/1711.01855}{{\tt
  arXiv:1711.01855 [hep-ph]}}].

\bibitem{Cepila:2016uku}
J.~Cepila, J.~G. Contreras and J.~D. Tapia~Takaki, {\it {Energy dependence of
  dissociative $\mathrm{J/}\psi$ photoproduction as a signature of gluon
  saturation at the LHC}},
  \href{http://dx.doi.org/10.1016/j.physletb.2016.12.063}{{\em Phys. Lett. B}
  {\bf 766} (2017) 186} [\href{http://arXiv.org/abs/1608.07559}{{\tt
  arXiv:1608.07559 [hep-ph]}}].

\bibitem{Kovchegov:2012mbw}
Y.~V. Kovchegov and E.~Levin, {\em {Quantum chromodynamics at high energy}},
  vol.~33.
\newblock Cambridge University Press, 8, 2012.

\bibitem{Iancu:2003xm}
E.~Iancu and R.~Venugopalan, {\it {The Color glass condensate and high-energy
  scattering in QCD}},
  \href{http://dx.doi.org/10.1142/9789812795533_0005}{{\em Quark gluon plasma
  3}  (2003) 249} [\href{http://arXiv.org/abs/hep-ph/0303204}{{\tt
  arXiv:hep-ph/0303204 [hep-ph]}}].
%%CITATION = HEP-PH/0303204;%%

\bibitem{Gelis:2010nm}
F.~Gelis, E.~Iancu, J.~Jalilian-Marian and R.~Venugopalan, {\it {The Color
  Glass Condensate}},
  \href{http://dx.doi.org/10.1146/annurev.nucl.010909.083629}{{\em Ann. Rev.
  Nucl. Part. Sci.} {\bf 60} (2010) 463}
  [\href{http://arXiv.org/abs/1002.0333}{{\tt arXiv:1002.0333 [hep-ph]}}].

\bibitem{Albacete:2014fwa}
J.~L. Albacete and C.~Marquet, {\it {Gluon saturation and initial conditions
  for relativistic heavy ion collisions}},
  \href{http://dx.doi.org/10.1016/j.ppnp.2014.01.004}{{\em Prog. Part. Nucl.
  Phys.} {\bf 76} (2014) 1} [\href{http://arXiv.org/abs/1401.4866}{{\tt
  arXiv:1401.4866 [hep-ph]}}].

\bibitem{Kowalski:2006hc}
H.~Kowalski, L.~Motyka and G.~Watt, {\it {Exclusive diffractive processes at
  HERA within the dipole picture}},
  \href{http://dx.doi.org/10.1103/PhysRevD.74.074016}{{\em Phys. Rev. D} {\bf
  74} (2006) 074016} [\href{http://arXiv.org/abs/hep-ph/0606272}{{\tt
  arXiv:hep-ph/0606272}}].

\bibitem{Hatta:2017cte}
Y.~Hatta, B.-W. Xiao and F.~Yuan, {\it {Gluon Tomography from Deeply Virtual
  Compton Scattering at Small-$x$}},
  \href{http://dx.doi.org/10.1103/PhysRevD.95.114026}{{\em Phys. Rev. D} {\bf
  95} (2017)~no.~11 114026} [\href{http://arXiv.org/abs/1703.02085}{{\tt
  arXiv:1703.02085 [hep-ph]}}].

\bibitem{Lappi:2020ufv}
T.~Lappi, H.~M\"antysaari and J.~Penttala, {\it {Relativistic corrections to
  the vector meson light front wave function}},
  \href{http://dx.doi.org/10.1103/PhysRevD.102.054020}{{\em Phys. Rev. D} {\bf
  102} (2020)~no.~5 054020} [\href{http://arXiv.org/abs/2006.02830}{{\tt
  arXiv:2006.02830 [hep-ph]}}].

\bibitem{Li:2017mlw}
Y.~Li, P.~Maris and J.~P. Vary, {\it {Quarkonium as a relativistic bound state
  on the light front}},
  \href{http://dx.doi.org/10.1103/PhysRevD.96.016022}{{\em Phys. Rev. D} {\bf
  96} (2017)~no.~1 016022} [\href{http://arXiv.org/abs/1704.06968}{{\tt
  arXiv:1704.06968 [hep-ph]}}].

\bibitem{Li:2021cwv}
M.~Li, Y.~Li, G.~Chen, T.~Lappi and J.~P. Vary, {\it {Light-front wavefunctions
  of mesons by design}},  \href{http://arXiv.org/abs/2111.07087}{{\tt
  arXiv:2111.07087 [hep-ph]}}.

\bibitem{Boussarie:2016bkq}
R.~Boussarie, A.~V. Grabovsky, D.~Y. Ivanov, L.~Szymanowski and S.~Wallon, {\it
  {Next-to-Leading Order Computation of Exclusive Diffractive Light Vector
  Meson Production in a Saturation Framework}},
  \href{http://dx.doi.org/10.1103/PhysRevLett.119.072002}{{\em Phys. Rev.
  Lett.} {\bf 119} (2017)~no.~7 072002}
  [\href{http://arXiv.org/abs/1612.08026}{{\tt arXiv:1612.08026 [hep-ph]}}].

\bibitem{Mantysaari:2021ryb}
H.~M\"antysaari and J.~Penttala, {\it {Exclusive heavy vector meson production
  at next-to-leading order in the dipole picture}},
  \href{http://dx.doi.org/10.1016/j.physletb.2021.136723}{{\em Phys. Lett. B}
  {\bf 823} (2021) 136723} [\href{http://arXiv.org/abs/2104.02349}{{\tt
  arXiv:2104.02349 [hep-ph]}}].

\bibitem{Beuf:2021srj}
G.~Beuf, T.~Lappi and R.~Paatelainen, {\it {Massive quarks at one loop in the
  dipole picture of Deep Inelastic Scattering}},
  \href{http://arXiv.org/abs/2112.03158}{{\tt arXiv:2112.03158 [hep-ph]}}.

\bibitem{Beuf:2020dxl}
G.~Beuf, H.~H\"anninen, T.~Lappi and H.~M\"antysaari, {\it {Color Glass
  Condensate at next-to-leading order meets HERA data}},
  \href{http://dx.doi.org/10.1103/PhysRevD.102.074028}{{\em Phys. Rev. D} {\bf
  102} (2020) 074028} [\href{http://arXiv.org/abs/2007.01645}{{\tt
  arXiv:2007.01645 [hep-ph]}}].

\bibitem{Beuf:2021qqa}
G.~Beuf, T.~Lappi and R.~Paatelainen, {\it {Massive quarks in NLO dipole
  factorization for DIS: Longitudinal photon}},
  \href{http://dx.doi.org/10.1103/PhysRevD.104.056032}{{\em Phys. Rev. D} {\bf
  104} (2021)~no.~5 056032} [\href{http://arXiv.org/abs/2103.14549}{{\tt
  arXiv:2103.14549 [hep-ph]}}].

\bibitem{Hanninen:2017ddy}
H.~H\"anninen, T.~Lappi and R.~Paatelainen, {\it {One-loop corrections to light
  cone wave functions: the dipole picture DIS cross section}},
  \href{http://dx.doi.org/10.1016/j.aop.2018.04.015}{{\em Annals Phys.} {\bf
  393} (2018) 358} [\href{http://arXiv.org/abs/1711.08207}{{\tt
  arXiv:1711.08207 [hep-ph]}}].

\bibitem{Beuf:2017bpd}
G.~Beuf, {\it {Dipole factorization for DIS at NLO: Combining the $q\bar{q}$
  and $q\bar{q}g$ contributions}},
  \href{http://dx.doi.org/10.1103/PhysRevD.96.074033}{{\em Phys. Rev. D} {\bf
  96} (2017)~no.~7 074033} [\href{http://arXiv.org/abs/1708.06557}{{\tt
  arXiv:1708.06557 [hep-ph]}}].

\bibitem{Beuf:2016wdz}
G.~Beuf, {\it {Dipole factorization for DIS at NLO: Loop correction to the
  $\gamma^*_{T,L}\to q\overline q$ light-front wave functions}},
  \href{http://dx.doi.org/10.1103/PhysRevD.94.054016}{{\em Phys. Rev. D} {\bf
  94} (2016)~no.~5 054016} [\href{http://arXiv.org/abs/1606.00777}{{\tt
  arXiv:1606.00777 [hep-ph]}}].

\bibitem{Lappi:2020srm}
T.~Lappi, H.~M\"antysaari and A.~Ramnath, {\it {Next-to-leading order
  Balitsky-Kovchegov equation beyond large $N_c$}},
  \href{http://dx.doi.org/10.1103/PhysRevD.102.074027}{{\em Phys. Rev. D} {\bf
  102} (2020)~no.~7 074027} [\href{http://arXiv.org/abs/2007.00751}{{\tt
  arXiv:2007.00751 [hep-ph]}}].

\bibitem{Lappi:2016fmu}
T.~Lappi and H.~M\"antysaari, {\it {Next-to-leading order Balitsky-Kovchegov
  equation with resummation}},
  \href{http://dx.doi.org/10.1103/PhysRevD.93.094004}{{\em Phys. Rev. D} {\bf
  93} (2016)~no.~9 094004} [\href{http://arXiv.org/abs/1601.06598}{{\tt
  arXiv:1601.06598 [hep-ph]}}].

\bibitem{Lappi:2015fma}
T.~Lappi and H.~M\"antysaari, {\it {Direct numerical solution of the coordinate
  space Balitsky-Kovchegov equation at next to leading order}},
  \href{http://dx.doi.org/10.1103/PhysRevD.91.074016}{{\em Phys. Rev. D} {\bf
  91} (2015)~no.~7 074016} [\href{http://arXiv.org/abs/1502.02400}{{\tt
  arXiv:1502.02400 [hep-ph]}}].

\bibitem{Balitsky:2008zza}
I.~Balitsky and G.~A. Chirilli, {\it {Next-to-leading order evolution of color
  dipoles}},  \href{http://dx.doi.org/10.1103/PhysRevD.77.014019}{{\em Phys.
  Rev. D} {\bf 77} (2008) 014019} [\href{http://arXiv.org/abs/0710.4330}{{\tt
  arXiv:0710.4330 [hep-ph]}}].

\bibitem{Ducloue:2019ezk}
B.~Duclou\'e, E.~Iancu, A.~H. Mueller, G.~Soyez and D.~N. Triantafyllopoulos,
  {\it {Non-linear evolution in QCD at high-energy beyond leading order}},
  \href{http://dx.doi.org/10.1007/JHEP04(2019)081}{{\em JHEP} {\bf 04} (2019)
  081} [\href{http://arXiv.org/abs/1902.06637}{{\tt arXiv:1902.06637
  [hep-ph]}}].

\bibitem{Ducloue:2019jmy}
B.~Duclou\'e, E.~Iancu, G.~Soyez and D.~N. Triantafyllopoulos, {\it {HERA data
  and collinearly-improved BK dynamics}},
  \href{http://dx.doi.org/10.1016/j.physletb.2020.135305}{{\em Phys. Lett. B}
  {\bf 803} (2020) 135305} [\href{http://arXiv.org/abs/1912.09196}{{\tt
  arXiv:1912.09196 [hep-ph]}}].

\bibitem{Iancu:2015vea}
E.~Iancu, J.~D. Madrigal, A.~H. Mueller, G.~Soyez and D.~N. Triantafyllopoulos,
  {\it {Resumming double logarithms in the QCD evolution of color dipoles}},
  \href{http://dx.doi.org/10.1016/j.physletb.2015.03.068}{{\em Phys. Lett. B}
  {\bf 744} (2015) 293} [\href{http://arXiv.org/abs/1502.05642}{{\tt
  arXiv:1502.05642 [hep-ph]}}].

\bibitem{Iancu:2015joa}
E.~Iancu, J.~D. Madrigal, A.~H. Mueller, G.~Soyez and D.~N. Triantafyllopoulos,
  {\it {Collinearly-improved BK evolution meets the HERA data}},
  \href{http://dx.doi.org/10.1016/j.physletb.2015.09.071}{{\em Phys. Lett. B}
  {\bf 750} (2015) 643} [\href{http://arXiv.org/abs/1507.03651}{{\tt
  arXiv:1507.03651 [hep-ph]}}].

\bibitem{Balitsky:2013fea}
I.~Balitsky and G.~A. Chirilli, {\it {Rapidity evolution of Wilson lines at the
  next-to-leading order}},
  \href{http://dx.doi.org/10.1103/PhysRevD.88.111501}{{\em Phys. Rev. D} {\bf
  88} (2013) 111501} [\href{http://arXiv.org/abs/1309.7644}{{\tt
  arXiv:1309.7644 [hep-ph]}}].

\bibitem{Kovner:2013ona}
A.~Kovner, M.~Lublinsky and Y.~Mulian, {\it {Jalilian-Marian, Iancu, McLerran,
  Weigert, Leonidov, Kovner evolution at next to leading order}},
  \href{http://dx.doi.org/10.1103/PhysRevD.89.061704}{{\em Phys. Rev. D} {\bf
  89} (2014)~no.~6 061704} [\href{http://arXiv.org/abs/1310.0378}{{\tt
  arXiv:1310.0378 [hep-ph]}}].

\bibitem{Caucal:2021ent}
P.~Caucal, F.~Salazar and R.~Venugopalan, {\it {Dijet impact factor in DIS at
  next-to-leading order in the Color Glass Condensate}},
  \href{http://dx.doi.org/10.1007/JHEP11(2021)222}{{\em JHEP} {\bf 11} (2021)
  222} [\href{http://arXiv.org/abs/2108.06347}{{\tt arXiv:2108.06347
  [hep-ph]}}].

\bibitem{Good:1960ba}
M.~L. Good and W.~D. Walker, {\it {Diffraction disssociation of beam
  particles}},  \href{http://dx.doi.org/10.1103/PhysRev.120.1857}{{\em Phys.
  Rev.} {\bf 120} (1960) 1857}.

\bibitem{Miettinen:1978jb}
H.~I. Miettinen and J.~Pumplin, {\it {Diffraction Scattering and the Parton
  Structure of Hadrons}},
  \href{http://dx.doi.org/10.1103/PhysRevD.18.1696}{{\em Phys. Rev. D} {\bf 18}
  (1978) 1696}.

\bibitem{Caldwell:2010zza}
A.~Caldwell and H.~Kowalski, {\it {Investigating the gluonic structure of
  nuclei via $\mathrm{J}/\psi$ scattering}},
  \href{http://dx.doi.org/10.1103/PhysRevC.81.025203}{{\em Phys. Rev. C} {\bf
  81} (2010) 025203} [\href{http://arXiv.org/abs/0909.1254}{{\tt
  arXiv:0909.1254}}].

\bibitem{Schenke:2012wb}
B.~Schenke, P.~Tribedy and R.~Venugopalan, {\it {Fluctuating Glasma initial
  conditions and flow in heavy ion collisions}},
  \href{http://dx.doi.org/10.1103/PhysRevLett.108.252301}{{\em Phys. Rev.
  Lett.} {\bf 108} (2012) 252301} [\href{http://arXiv.org/abs/1202.6646}{{\tt
  arXiv:1202.6646 [nucl-th]}}].

\bibitem{Mantysaari:2020lhf}
H.~M\"antysaari, K.~Roy, F.~Salazar and B.~Schenke, {\it {Gluon imaging using
  azimuthal correlations in diffractive scattering at the Electron-Ion
  Collider}},  \href{http://dx.doi.org/10.1103/PhysRevD.103.094026}{{\em Phys.
  Rev. D} {\bf 103} (2021)~no.~9 094026}
  [\href{http://arXiv.org/abs/2011.02464}{{\tt arXiv:2011.02464 [hep-ph]}}].

\bibitem{Mantysaari:2019jhh}
H.~M\"antysaari and B.~Schenke, {\it {Accessing the gluonic structure of light
  nuclei at a future electron-ion collider}},
  \href{http://dx.doi.org/10.1103/PhysRevC.101.015203}{{\em Phys. Rev. C} {\bf
  101} (2020)~no.~1 015203} [\href{http://arXiv.org/abs/1910.03297}{{\tt
  arXiv:1910.03297 [hep-ph]}}].

\bibitem{Mantysaari:2019csc}
H.~M\"antysaari, N.~Mueller and B.~Schenke, {\it {Diffractive Dijet Production
  and Wigner Distributions from the Color Glass Condensate}},
  \href{http://dx.doi.org/10.1103/PhysRevD.99.074004}{{\em Phys. Rev. D} {\bf
  99} (2019)~no.~7 074004} [\href{http://arXiv.org/abs/1902.05087}{{\tt
  arXiv:1902.05087 [hep-ph]}}].

\bibitem{Mantysaari:2018zdd}
H.~M\"antysaari and B.~Schenke, {\it {Confronting impact parameter dependent
  JIMWLK evolution with HERA data}},
  \href{http://dx.doi.org/10.1103/PhysRevD.98.034013}{{\em Phys. Rev. D} {\bf
  98} (2018)~no.~3 034013} [\href{http://arXiv.org/abs/1806.06783}{{\tt
  arXiv:1806.06783 [hep-ph]}}].

\bibitem{McLerran:1993ni}
L.~D. McLerran and R.~Venugopalan, {\it {Computing quark and gluon distribution
  functions for very large nuclei}},
  \href{http://dx.doi.org/10.1103/PhysRevD.49.2233}{{\em Phys. Rev. D} {\bf 49}
  (1994) 2233} [\href{http://arXiv.org/abs/hep-ph/9309289}{{\tt
  arXiv:hep-ph/9309289}}].

\bibitem{Kowalski:2003hm}
H.~Kowalski and D.~Teaney, {\it {An Impact parameter dipole saturation model}},
   \href{http://dx.doi.org/10.1103/PhysRevD.68.114005}{{\em Phys. Rev. D} {\bf
  68} (2003) 114005} [\href{http://arXiv.org/abs/hep-ph/0304189}{{\tt
  arXiv:hep-ph/0304189}}].

\bibitem{Rezaeian:2012ji}
A.~H. Rezaeian, M.~Siddikov, M.~Van~de Klundert and R.~Venugopalan, {\it
  {Analysis of combined HERA data in the Impact-Parameter dependent Saturation
  model}},  \href{http://dx.doi.org/10.1103/PhysRevD.87.034002}{{\em Phys. Rev.
  D} {\bf 87} (2013)~no.~3 034002} [\href{http://arXiv.org/abs/1212.2974}{{\tt
  arXiv:1212.2974 [hep-ph]}}].

\bibitem{Mantysaari:2018nng}
H.~M\"antysaari and P.~Zurita, {\it {In depth analysis of the combined HERA
  data in the dipole models with and without saturation}},
  \href{http://dx.doi.org/10.1103/PhysRevD.98.036002}{{\em Phys. Rev. D} {\bf
  98} (2018) 036002} [\href{http://arXiv.org/abs/1804.05311}{{\tt
  arXiv:1804.05311 [hep-ph]}}].

\bibitem{Lappi:2007ku}
T.~Lappi, {\it {Wilson line correlator in the MV model: Relating the glasma to
  deep inelastic scattering}},
  \href{http://dx.doi.org/10.1140/epjc/s10052-008-0588-4}{{\em Eur. Phys. J. C}
  {\bf 55} (2008) 285} [\href{http://arXiv.org/abs/0711.3039}{{\tt
  arXiv:0711.3039 [hep-ph]}}].

\bibitem{Albacete:2016pmp}
J.~L. Albacete and A.~Soto-Ontoso, {\it {Hot spots and the hollowness of
  proton\textendash{}proton interactions at high energies}},
  \href{http://dx.doi.org/10.1016/j.physletb.2017.04.055}{{\em Phys. Lett. B}
  {\bf 770} (2017) 149} [\href{http://arXiv.org/abs/1605.09176}{{\tt
  arXiv:1605.09176 [hep-ph]}}].

\bibitem{Albacete:2017ajt}
J.~L. Albacete, H.~Petersen and A.~Soto-Ontoso, {\it {Symmetric cumulants as a
  probe of the proton substructure at LHC energies}},
  \href{http://dx.doi.org/10.1016/j.physletb.2018.01.011}{{\em Phys. Lett. B}
  {\bf 778} (2018) 128} [\href{http://arXiv.org/abs/1707.05592}{{\tt
  arXiv:1707.05592 [hep-ph]}}].

\bibitem{Alkin:2014rfa}
A.~Alkin, E.~Martynov, O.~Kovalenko and S.~M. Troshin, {\it {Impact-parameter
  analysis of TOTEM data at the LHC: Black disk limit exceeded}},
  \href{http://dx.doi.org/10.1103/PhysRevD.89.091501}{{\em Phys. Rev. D} {\bf
  89} (2014)~no.~9 091501} [\href{http://arXiv.org/abs/1403.8036}{{\tt
  arXiv:1403.8036 [hep-ph]}}].

\bibitem{Dremin:2015ujt}
I.~M. Dremin, {\it {Will protons become gray at 13 TeV and 100 TeV?}},
  \href{http://dx.doi.org/10.3103/S1068335617040029}{{\em Bull. Lebedev Phys.
  Inst.} {\bf 44} (2017)~no.~4 94} [\href{http://arXiv.org/abs/1511.03212}{{\tt
  arXiv:1511.03212 [hep-ph]}}].

\bibitem{Troshin:2016frs}
S.~M. Troshin and N.~E. Tyurin, {\it {The new scattering mode emerging at the
  LHC?}},  \href{http://dx.doi.org/10.1142/S0217732316500796}{{\em Mod. Phys.
  Lett. A} {\bf 31} (2016)~no.~13 1650079}
  [\href{http://arXiv.org/abs/1602.08972}{{\tt arXiv:1602.08972 [hep-ph]}}].

\bibitem{Arriola:2016bxa}
E.~Ruiz~Arriola and W.~Broniowski, {\it {Proton\textendash{}Proton On Shell
  Optical Potential at High Energies and the Hollowness Effect}},
  \href{http://dx.doi.org/10.1007/s00601-016-1095-z}{{\em Few Body Syst.} {\bf
  57} (2016)~no.~7 485} [\href{http://arXiv.org/abs/1602.00288}{{\tt
  arXiv:1602.00288 [hep-ph]}}].

\bibitem{Sirunyan:2017uyl}
{\bf CMS} collaboration, A.~M. Sirunyan {\em et.~al.}, {\it {Observation of
  Correlated Azimuthal Anisotropy Fourier Harmonics in $pp$ and $p+Pb$
  Collisions at the LHC}},
  \href{http://dx.doi.org/10.1103/PhysRevLett.120.092301}{{\em Phys. Rev.
  Lett.} {\bf 120} (2018)~no.~9 092301}
  [\href{http://arXiv.org/abs/1709.09189}{{\tt arXiv:1709.09189 [nucl-ex]}}].

\bibitem{Acharya:2019vdf}
{\bf ALICE} collaboration, S.~Acharya {\em et.~al.}, {\it {Investigations of
  Anisotropic Flow Using Multiparticle Azimuthal Correlations in pp, p-Pb,
  Xe-Xe, and Pb-Pb Collisions at the LHC}},
  \href{http://dx.doi.org/10.1103/PhysRevLett.123.142301}{{\em Phys. Rev.
  Lett.} {\bf 123} (2019)~no.~14 142301}
  [\href{http://arXiv.org/abs/1903.01790}{{\tt arXiv:1903.01790 [nucl-ex]}}].

\bibitem{Moreland:2014oya}
J.~S. Moreland, J.~E. Bernhard and S.~A. Bass, {\it {Alternative ansatz to
  wounded nucleon and binary collision scaling in high-energy nuclear
  collisions}},  \href{http://dx.doi.org/10.1103/PhysRevC.92.011901}{{\em Phys.
  Rev. C} {\bf 92} (2015)~no.~1 011901}
  [\href{http://arXiv.org/abs/1412.4708}{{\tt arXiv:1412.4708 [nucl-th]}}].

\bibitem{McLerran:2015qxa}
L.~McLerran and P.~Tribedy, {\it {Intrinsic Fluctuations of the Proton
  Saturation Momentum Scale in High Multiplicity p+p Collisions}},
  \href{http://dx.doi.org/10.1016/j.nuclphysa.2015.10.008}{{\em Nucl. Phys. A}
  {\bf 945} (2016) 216} [\href{http://arXiv.org/abs/1508.03292}{{\tt
  arXiv:1508.03292 [hep-ph]}}].

\bibitem{sivia2006data}
D.~Sivia and J.~Skilling, {\em Data analysis: a Bayesian tutorial}.
\newblock OUP Oxford, 2006.

\bibitem{williams2006gaussian}
C.~K. Williams and C.~E. Rasmussen, {\em Gaussian processes for machine
  learning}, vol.~2.
\newblock MIT press Cambridge, MA, 2006.

\bibitem{Bernhard:2015hxa}
J.~E. Bernhard, P.~W. Marcy, C.~E. Coleman-Smith, S.~Huzurbazar, R.~L. Wolpert
  and S.~A. Bass, {\it {Quantifying properties of hot and dense QCD matter
  through systematic model-to-data comparison}},
  \href{http://dx.doi.org/10.1103/PhysRevC.91.054910}{{\em Phys. Rev. C} {\bf
  91} (2015)~no.~5 054910} [\href{http://arXiv.org/abs/1502.00339}{{\tt
  arXiv:1502.00339 [nucl-th]}}].

\bibitem{goodman2010ensemble}
J.~Goodman and J.~Weare, {\it Ensemble samplers with affine invariance},  {\em
  Communications in applied mathematics and computational science} {\bf 5}
  (2010)~no.~1 65.

\bibitem{foreman2013emcee}
D.~Foreman-Mackey, D.~W. Hogg, D.~Lang and J.~Goodman, {\it emcee: the mcmc
  hammer},  {\em Publications of the Astronomical Society of the Pacific} {\bf
  125} (2013)~no.~925 306.

\bibitem{Nijs:2021clz}
G.~Nijs and W.~van~der Schee, {\it {Predictions and postdictions for
  relativistic lead and oxygen collisions with $Trajectum$}},
  \href{http://arXiv.org/abs/2110.13153}{{\tt arXiv:2110.13153 [nucl-th]}}.

\bibitem{Gribov:1972ri}
V.~N. Gribov and L.~N. Lipatov, {\it {Deep inelastic e p scattering in
  perturbation theory}},  {\em Sov. J. Nucl. Phys.} {\bf 15} (1972) 438.

\bibitem{Gribov:1972rt}
V.~N. Gribov and L.~N. Lipatov, {\it {$e^+ e^-$ pair annihilation and deep
  inelastic e p scattering in perturbation theory}},  {\em Sov. J. Nucl. Phys.}
  {\bf 15} (1972) 675.

\bibitem{Altarelli:1977zs}
G.~Altarelli and G.~Parisi, {\it {Asymptotic Freedom in Parton Language}},
  \href{http://dx.doi.org/10.1016/0550-3213(77)90384-4}{{\em Nucl. Phys. B}
  {\bf 126} (1977) 298}.

\bibitem{Dokshitzer:1977sg}
Y.~L. Dokshitzer, {\it {Calculation of the Structure Functions for Deep
  Inelastic Scattering and $e^+ e^-$ Annihilation by Perturbation Theory in
  Quantum Chromodynamics.}},  {\em Sov. Phys. JETP} {\bf 46} (1977) 641.

\bibitem{Lappi:2010dd}
T.~Lappi and H.~Mäntysaari, {\it {Incoherent diffractive $\mathrm{J}/\psi$
  production in high energy nuclear DIS}},
  \href{http://dx.doi.org/10.1103/PhysRevC.83.065202}{{\em Phys. Rev. C} {\bf
  83} (2011) 065202} [\href{http://arXiv.org/abs/1011.1988}{{\tt
  arXiv:1011.1988 [hep-ph]}}].

\bibitem{Jalilian-Marian:1996mkd}
J.~Jalilian-Marian, A.~Kovner, L.~D. McLerran and H.~Weigert, {\it {The
  Intrinsic glue distribution at very small $x$}},
  \href{http://dx.doi.org/10.1103/PhysRevD.55.5414}{{\em Phys. Rev. D} {\bf 55}
  (1997) 5414} [\href{http://arXiv.org/abs/hep-ph/9606337}{{\tt
  arXiv:hep-ph/9606337}}].

\bibitem{Jalilian-Marian:1997qno}
J.~Jalilian-Marian, A.~Kovner, A.~Leonidov and H.~Weigert, {\it {The BFKL
  equation from the Wilson renormalization group}},
  \href{http://dx.doi.org/10.1016/S0550-3213(97)00440-9}{{\em Nucl. Phys. B}
  {\bf 504} (1997) 415} [\href{http://arXiv.org/abs/hep-ph/9701284}{{\tt
  arXiv:hep-ph/9701284}}].

\bibitem{Jalilian-Marian:1997jhx}
J.~Jalilian-Marian, A.~Kovner, A.~Leonidov and H.~Weigert, {\it {The Wilson
  renormalization group for low x physics: Towards the high density regime}},
  \href{http://dx.doi.org/10.1103/PhysRevD.59.014014}{{\em Phys. Rev. D} {\bf
  59} (1998) 014014} [\href{http://arXiv.org/abs/hep-ph/9706377}{{\tt
  arXiv:hep-ph/9706377}}].

\bibitem{Iancu:2001md}
E.~Iancu and L.~D. McLerran, {\it {Saturation and universality in QCD at small
  $x$}},  \href{http://dx.doi.org/10.1016/S0370-2693(01)00526-3}{{\em Phys.
  Lett. B} {\bf 510} (2001) 145}
  [\href{http://arXiv.org/abs/hep-ph/0103032}{{\tt arXiv:hep-ph/0103032}}].

\bibitem{Ferreiro:2001qy}
E.~Ferreiro, E.~Iancu, A.~Leonidov and L.~McLerran, {\it {Nonlinear gluon
  evolution in the color glass condensate. 2.}},
  \href{http://dx.doi.org/10.1016/S0375-9474(01)01329-X}{{\em Nucl. Phys. A}
  {\bf 703} (2002) 489} [\href{http://arXiv.org/abs/hep-ph/0109115}{{\tt
  arXiv:hep-ph/0109115}}].

\bibitem{Iancu:2001ad}
E.~Iancu, A.~Leonidov and L.~D. McLerran, {\it {The Renormalization group
  equation for the color glass condensate}},
  \href{http://dx.doi.org/10.1016/S0370-2693(01)00524-X}{{\em Phys. Lett. B}
  {\bf 510} (2001) 133} [\href{http://arXiv.org/abs/hep-ph/0102009}{{\tt
  arXiv:hep-ph/0102009}}].

\bibitem{Iancu:2000hn}
E.~Iancu, A.~Leonidov and L.~D. McLerran, {\it {Nonlinear gluon evolution in
  the color glass condensate. 1.}},
  \href{http://dx.doi.org/10.1016/S0375-9474(01)00642-X}{{\em Nucl. Phys. A}
  {\bf 692} (2001) 583} [\href{http://arXiv.org/abs/hep-ph/0011241}{{\tt
  arXiv:hep-ph/0011241}}].

\bibitem{Mueller:2001uk}
A.~H. Mueller, {\it {A Simple derivation of the JIMWLK equation}},
  \href{http://dx.doi.org/10.1016/S0370-2693(01)01343-0}{{\em Phys. Lett. B}
  {\bf 523} (2001) 243} [\href{http://arXiv.org/abs/hep-ph/0110169}{{\tt
  arXiv:hep-ph/0110169}}].

\bibitem{ZEUS:2002wfj}
{\bf ZEUS} collaboration, S.~Chekanov {\em et.~al.}, {\it {Exclusive
  photoproduction of $\mathrm{J}/\psi$ mesons at HERA}},
  \href{http://dx.doi.org/10.1007/s10052-002-0953-7}{{\em Eur. Phys. J. C} {\bf
  24} (2002) 345} [\href{http://arXiv.org/abs/hep-ex/0201043}{{\tt
  arXiv:hep-ex/0201043}}].

\bibitem{Schlichting:2014ipa}
S.~Schlichting and B.~Schenke, {\it {The shape of the proton at high
  energies}},  \href{http://dx.doi.org/10.1016/j.physletb.2014.10.068}{{\em
  Phys. Lett. B} {\bf 739} (2014) 313}
  [\href{http://arXiv.org/abs/1407.8458}{{\tt arXiv:1407.8458 [hep-ph]}}].

\bibitem{H1:2005dtp}
{\bf H1} collaboration, A.~Aktas {\em et.~al.}, {\it {Elastic $\mathrm{J}/\psi$
  production at HERA}},
  \href{http://dx.doi.org/10.1140/epjc/s2006-02519-5}{{\em Eur. Phys. J. C}
  {\bf 46} (2006) 585} [\href{http://arXiv.org/abs/hep-ex/0510016}{{\tt
  arXiv:hep-ex/0510016}}].

\bibitem{Aaij:2014iea}
{\bf LHCb} collaboration, R.~Aaij {\em et.~al.}, {\it {Updated measurements of
  exclusive $\mathrm{J}/\psi$ and $\psi$(2S) production cross-sections in pp
  collisions at $\sqrt{s}=7$ TeV}},
  \href{http://dx.doi.org/10.1088/0954-3899/41/5/055002}{{\em J. Phys. G} {\bf
  41} (2014) 055002} [\href{http://arXiv.org/abs/1401.3288}{{\tt
  arXiv:1401.3288 [hep-ex]}}].

\bibitem{Acharya:2018jua}
{\bf ALICE} collaboration, S.~Acharya {\em et.~al.}, {\it {Energy dependence of
  exclusive $\mathrm {J}/\psi $ photoproduction off protons in ultra-peripheral
  p\textendash{}Pb collisions at $\sqrt{s_{\mathrm {\scriptscriptstyle NN}}} =
  5.02$ TeV}},  \href{http://dx.doi.org/10.1140/epjc/s10052-019-6816-2}{{\em
  Eur. Phys. J. C} {\bf 79} (2019)~no.~5 402}
  [\href{http://arXiv.org/abs/1809.03235}{{\tt arXiv:1809.03235 [nucl-ex]}}].

\bibitem{Mantysaari:2017cni}
H.~M\"antysaari, B.~Schenke, C.~Shen and P.~Tribedy, {\it {Imprints of
  fluctuating proton shapes on flow in proton-lead collisions at the LHC}},
  \href{http://dx.doi.org/10.1016/j.physletb.2017.07.038}{{\em Phys. Lett. B}
  {\bf 772} (2017) 681} [\href{http://arXiv.org/abs/1705.03177}{{\tt
  arXiv:1705.03177 [nucl-th]}}].

\bibitem{IPGlasmaDiffraction}
The overarching framework for IPGlasma + subnucleonic diffraction model is
  available at \url{https://github.com/chunshen1987/IPGlasmaFramework}
  (v1.0.0). It uses the open-source code packages IP-Glasma
  \url{https://github.com/schenke/ipglasma} and the program to compute
  exclusive J/$\psi$ production cross-section
  \url{https://github.com/hejajama/subnucleondiffraction}.

\bibitem{BayesianPackage}
The numerical package for our Bayesian analysis is available at
  \url{https://github.com/chunshen1987/bayesian_analysis/releases/tag/v1.0.0}.

\bibitem{StreamlitApp}
An interactive emulator for our model is available at
  \url{https://share.streamlit.io/chunshen1987/ipglasmadiffractionstreamlit/main/IPGlasmaDiffraction_app.py}.

\bibitem{Pordes:2007zzb}
R.~Pordes {\em et.~al.}, {\it {The Open Science Grid}},
  \href{http://dx.doi.org/10.1088/1742-6596/78/1/012057}{{\em J. Phys. Conf.
  Ser.} {\bf 78} (2007) 012057}.

\bibitem{Sfiligoi:2009cct}
I.~Sfiligoi, D.~C. Bradley, B.~Holzman, P.~Mhashilkar, S.~Padhi and
  F.~Wurthwrin, {\it {The pilot way to Grid resources using glideinWMS}},
  \href{http://dx.doi.org/10.1109/CSIE.2009.950}{{\em WRI World Congress} {\bf
  2} (2009) 428}.

\end{thebibliography}\endgroup

\end{document}